\documentclass[conference]{IEEEtran}

\usepackage[utf8]{inputenc}
\usepackage{graphicx}
\usepackage{etex}
\usepackage{amsfonts}
\usepackage{amssymb}
\usepackage{url}
\usepackage{paralist}
\usepackage{graphtex}
\usepackage{listings}
\usepackage[tight, footnotesize]{subfigure}
\usepackage[all]{xy}
\usepackage{color}
\usepackage{ifpdf}

\ifCLASSINFOpdf
  \usepackage{graphicx}
\else
  \usepackage[dvips]{graphicx}
\fi

\usepackage[cmex10]{amsmath}
\renewcommand{\baselinestretch}{0.98}

\newcommand{\Sa}{\Sigma}

\newcommand{\e}{\varepsilon}

\lstdefinelanguage{Algo}{ morekeywords = {do, let, if, then, else, elseif, fi, while, endwhile, for, return, function, done, new, end, switch, case, in, instanceof},
  sensitive, %
  morecomment = [l]\#, %
  morecomment = [l]//, %
  morecomment = [s]{/*}{*/}, %
  morestring = [b]'', %
  morestring = [b]'%
}[keywords, comments, strings]

\begin{document}

\title{Controlled~conflict~resolution~for~replicated~document}
%Format\titlenote{(Does NOT produce the permission block, copyright information nor page numbering). For use with ACM\_PROC\_ARTICLE-SP.CLS. Supported by ACM.}}
%\subtitle{[Extended Abstract]
%\titlenote{A full version of this paper is available as
%\textit{Author's Guide to Preparing ACM SIG Proceedings Using
%\LaTeX$2_\epsilon$\ and BibTeX} at
%\texttt{www.acm.org/eaddress.htm}}}

\author{
% You can go ahead and credit any number of authors here,
% e.g. one 'row of three' or two rows (consisting of one row of three
% and a second row of one, two or three).
%
% The command \alignauthor (no curly braces needed) should
% precede each author name, affiliation/snail-mail address and
% e-mail address. Additionally, tag each line of
% affiliation/address with \affaddr, and tag the
% e-mail address with \email.
%
\IEEEauthorblockN{Stéphane Martin} %\titlenote{The secretary disavows
%ny knowledge of this author's actions.}\\
\IEEEauthorblockA{Université de Lorraine\\
CNRS -- INRIA -- LORIA\\%Campus Scientifique - BP 239, \\
54500 Vandoeuvre-lès-Nancy, France\\
Email: stephane.martin@loria.fr }
  \and
\IEEEauthorblockN{Mehdi Ahmed-Nacer}%\titlenote{This author is the
\IEEEauthorblockA{Université de Lorraine\\
CNRS -- INRIA -- LORIA\\
%Campus Scientifique - BP 239, \\
54500 Vandoeuvre-lès-Nancy, France\\
Email: mehdi.ahmed-nacer@loria.fr}

\and
\IEEEauthorblockN{Pascal Urso}%\titlenote{This author is the
\IEEEauthorblockA{Université de Lorraine\\
CNRS -- INRIA -- LORIA\\
%Campus Scientifique - BP 239, \\
54500 Vandoeuvre-lès-Nancy, France\\
Email: pascal.urso@loria.fr}
}

\maketitle

\begin{abstract} 
  Collaborative working is increasingly popular, but it presents
  challenges due to the need for high responsiveness and disconnected
  work support. To address these challenges the data is optimistically
  replicated at the edges of the network, i.e. personal computers or
  mobile devices. This replication requires a merge mechanism that
  preserves the consistency and structure of the shared data 
  subject to concurrent modifications.

  In this paper, we propose a generic design to ensure eventual
  consistency (every replica will eventually view the same data) and to maintain 
  the specific constraints of the replicated data. Our layered
  design provides to the application engineer the complete control
  over system scalability and behavior of the replicated data in face
  of concurrent modifications. We show that our design allows
  replication of complex data types with acceptable performances.
\end{abstract}

%% A category with the (minimum) three required fields
%\category{I.7.1}{Document and Text Processing}{Document and Text Editing}
%\category{C.2.4}{Computer-Communication Networks}{Distributed Systems}[Distributed applications]
%\category{E.1}{Data}{Data Structures}[Trees]
%%A category including the fourth, optional field follows...
%\category{D.2.2}{Software Engineering}{Design Tools and Techniques}[Structured programming]

%\terms{Design}

% NOT required for Proceedings

%--------------------------------------------------------------
% Comment 
% 
% 

%\section{Introduction}
%
\IEEEpeerreviewmaketitle

%\section{Introduction}
%qu'est ce que la réplication optimiste ?
\section{Introduction}
Replication allows accessibility of shared data in collaborative tools
(such as Google Docs) and mobile applications (such as Evernote or
Dropbox). Indeed, collaboration is achieved by different distinct
sites that work independently on a replica, i.e. a copy of the
document. Due to high responsiveness and disconnected work
requirements, such applications cannot use lock or consensus
mechanisms. %check

However, the CAP theorem~\cite{gilbert02brewer} states that a
replicated system cannot ensure strong Consistency together with
Availability and Partition tolerance. In such applications, where
availability is required by users and partition is unavoidable, a
solution is temporal divergence of replicas, i.e. to use
optimistic replication. Of course, at the end of the modification process,
users aim to have the same document. This kind of consistency model is
called ``{\em eventual consistency}'' which guarantees that if no new
update is made to the object, eventually all accesses will return
the same value. To obtain eventual consistency, a particular merge
procedure that handles conflicting concurrent
modifications, is required.

We consider that two concurrent modifications {\em conflict}, if, once
both integrated, they violate the structural constraints of a data
type. For instance, with a replicated structured document, adding
concurrently two titles conflicts if the document type accepts only one
title. To obtain a conflict-free replicated data type, the merge
procedure must make an arbitrary choice (such as: appending the
titles, ``priority-replica-wins'', ``last-writer-wins'', etc.). Moreover, every
replica must make independently the same choice. Conflict resolution
is also a question of scalability and performances since different
choice procedures may have different computational complexities.

Unfortunately, eventual consistency is more difficult to achieve
facing complex conflict resolution as demonstrated by the numerous
proposed approaches that fail to ensure it for simple plain text
document~\cite{imine03proving, oster06tombstone}. Indeed, more the data
type is complex, more conflicts appear. For instance, in a
hierarchical document, modifications such as adding and removing an
element, or adding a paragraph while removing the section to which it
belongs, or setting concurrently two titles conflict.

We propose a framework that decouples eventual consistency management from data
type constraints satisfaction. Our framework is made of layers.  A layer can
use the result of one or more independent layers.  The lowest layer hosts the
replicated data structure and are in charge to merge concurrent modifications.
These lowest layers encapsulate an existing eventually consistent data type
from the literature. Other layers are in charge to ensure a constraint on a
data type. It does not modify the inner state of the replicated data but only
computes a view that satisfies the constraint. 

Our framework manages each conflict type independently while assuring
eventual consistency. Thanks to layered design, any combination of
conflict resolution is designable, giving to the application the
entire control on the system scalability and behavior of the
replicated data in face of concurrent mutations.

\section{Motivation}

Our approach is based on the observation that obtaining eventual
consistency while ensuring complex constraints on a data type is
difficult. Thus, we propose to decouple eventual consistency from data
integrity insurance trough layers. 

To illustrate the behavior of such a decoupling, let's imagine a
replicated file system. Ensuring eventual consistency of a file system
is complex~\cite{richard1992ficus}, while ensuring eventual
consistency of a set can be achieved in numerous ways with quite simple
algorithms. For instance, \cite{shapiro11comprehensive} defines
multiple replicated sets with different behaviors and performances.

So we can imagine a file system as the set of absolute paths present
in the file system.
\begin{enumerate} 
\item A first layer contains the set of independent couples $(path,
  type)$ which are elements present in the file system. Types can be
  directory or file. This layer communicates with the first layer of
  the other replicas. It transmits simple messages that correspond to
  an addition or a suppression in the set. This layer ensures alone
  eventual consistency by merging these messages.
\item The second layer is in charge of producing a tree from the set
  of paths. To produce this tree, it must ensure the constraint that
  all nodes are accessible by the root. Indeed if a replica removes a
  directory, while another adds a file into this directory, the path
  to the file is present in the set while the path to the directory is
  not. Such a layer may drop this ``orphan'' file or place it under
  some special ``lost-and-found'' directory (see
  Section~\ref{sec:unordered}).
\item The third layer is in charge of producing a file system from the
  tree. It satisfies the unique name constraint on a directory. Indeed,
  a directory may contains two children (one directory and one file)
  added concurrently with the same name. Such a layer may rename
  elements, or enforce specific name when adding an element (files and
  only files must have an extension, such as \verb|.java|).
\end{enumerate}

Replicated file systems (and some other complex data types), already exist in
the literature. The advantage of our model is twofold. The first advantage is
that only the first layer is in charge of merging concurrent operations. For
the other layers, the data is handled as local data, simplifying the eventual
consistency issues. The second advantage is the modularity of the approach. A
layer that provides a data type can be freely substituted by another
implementation. Thus, our approach can provide many different behaviors, while
each existing solution proposes only one or a small number of different
behavior(s) with an associated performance level which could not be appropriate
to every collaborative application context.

\section{Layered data types}

We define a data type as an object with a two methods interface: i)
the ''lookup`` method returns the data type state; ii) the ''modify``
method performs modifications in the data type state.

A {\em replicated data type} is a data type with a communication
interface to merge its state with other replicas. Concretely, on each
update invocation from an application, the replicated data type sends
to another replica a message that represents the local modification. A
replicated data type which receives such a message, integrates it on
its own state. We require that a replicated data type ensures eventual
consistency. This means that, after all modifications were performed,
the invocation of the lookup method eventually returns the same result.

First, we encapsulate an existing eventually consistent data type in a
{\em replication layer}. This kind of layer is the bottom layer of
our model. It ensures communication between replicas and manages
concurrent modifications. The other kind of layer we define is the {\em
  adaptation layer} that uses the data provided by one or more layers
and ensures a particular constraint on the data type. An adaptation
layer can be placed on top of one or more layers that can be
replication or adaptation layers.

\begin{figure}[htb]
{\centering
\includegraphics[width = 8.5cm]{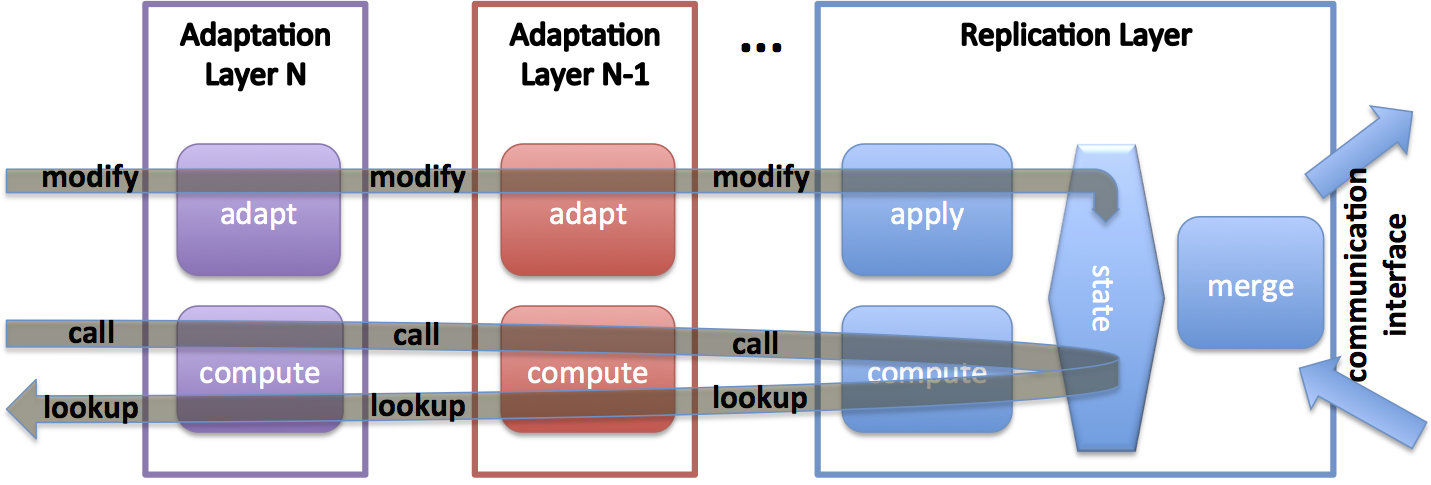}}
\caption{Layers}
\label{fig:dynamic}
\end{figure}

As presented in Figure~\ref{fig:dynamic}, the generic computational
aspect of our model is quite simple. When an application modifies a data
type, it calls the higher layer modify function. The higher layer
adapts the given local operation into one or more local operation(s)
applied on the layer just below. This layer will itself adapt these
local operations for the third layer, and so on until the replication
layer. Only the replication layer is in charge to communicate local
updates to other replicas and to merge local and remote modifications.
When the application asks for the value of the data type, it calls the
higher layer lookup interface. The layer calls the lookup interface of
the layer just below and computes a result corresponding to the
application needs.

The lookup method of an adaptation layer recomputes totally its result
from the inner layer(s) lookup invocation(s) result(s). {\em This
  computation does not affect the inner-layer state, if any.} Assuming
this computation is deterministic and that the below layer(s) ensure(s)
eventual consistency, we can prove straight-forwardly that the
adaptation layer provides an eventually consistent data type.

Such a computation must be done when a view is requested, but only if
the inner data was modified since the last request. This is adapted to
{\em state-based} replication mechanisms~\cite{saito05optimistic}
(such as version control systems). State-based replication mechanisms
transfer their whole state to other replicas, thus, fewer merge occurs
but each merge may modify up to the whole state of the data.

However, for {\em operation-based} replication
mechanisms~\cite{saito05optimistic}, we should define incremental
adaptation layers. Operation-based replication mechanisms sends update
operations (or differences).

\subsection*{Incremental Layers}

\begin{figure}[htb]
{\centering
\includegraphics[width = 8.5cm]{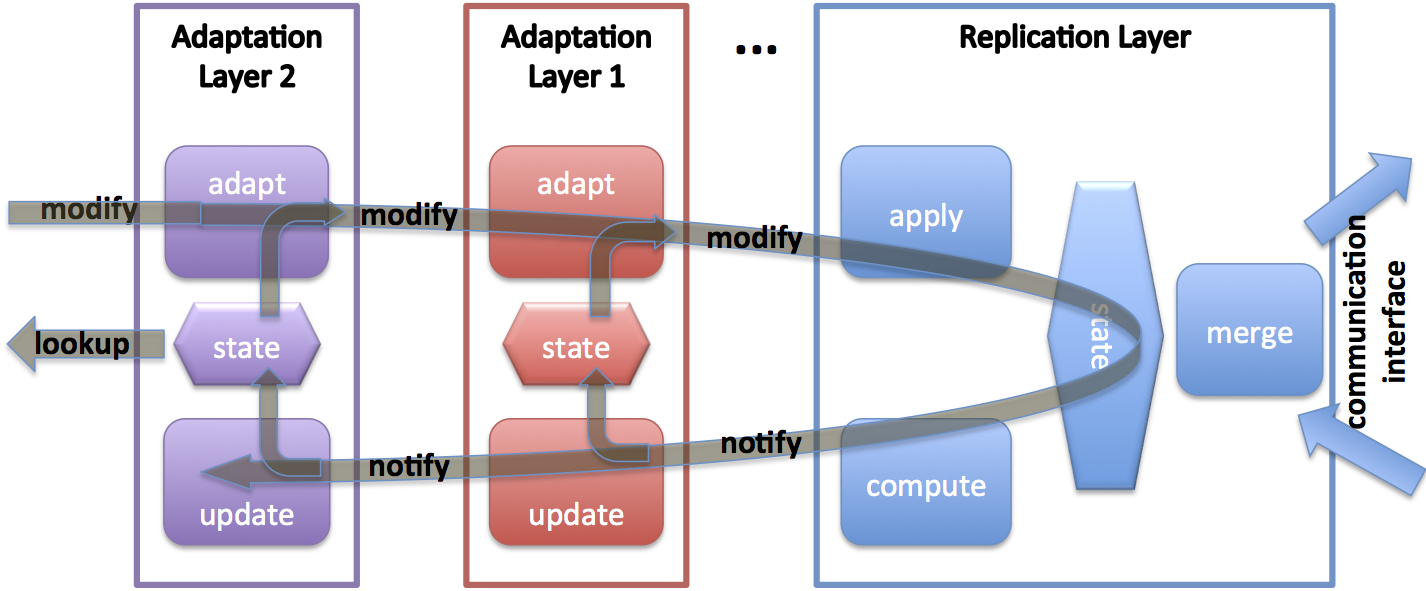}}
\caption{Incremental layers}
\label{fig:incremental}
\end{figure}

An incremental adaptation layer stores the state of the data type that
will be returned to the application. It modifies this data type each
time its inner layer state is modified, following an observer design
pattern, see Figure~\ref{fig:incremental}. Therefore, it modifies only
a part of the data type.  Potentially, an incremental lookup has
better performances. Eventual consistency can be ensured by an
equivalence between the incremental lookup and some non-incremental
lookup. Anyway, as non-incremental layers, incremental layers
computations do not affect their inner-layer state.

Even if incremental layers seem more adapted to operation-based replication
mechanisms, any combination of layers can be constructed.  Indeed, a
state-based replication layer that notifies changes to its observers can be
used below an incremental layer. Also, an incremental layer can be used below
a non-incremental one.\footnote{This last combination can be useful when no
incremental solution is available for a given constraint (for XSD schema
repairing for instance).}

% \begin{theorem} A layered data type is eventually consistent if the
%   result of the lookup method is equivalent to a deterministic
%   function applied to the inner lookup result and if the inner
%   data type ensures eventual consistency.
% \end{theorem}

\section{Examples}
\label{sec:examples}

This section presents several examples of data types that can be
obtained using our framework. Due to space limitation, only some of
them will be completely detailed.

\subsection{Text data type}
\label{order}

In this section, we show how to obtain a text data type, i.e. an
ordered sequence of elements (lines, character, or paragraphs,
etc.). Beside its apparent simplicity, this a non-trivial problem as
evidenced by the huge literature on the
subject:~\cite{oster06tombstone, weiss07wooki, preguica09commutative}.
The challenge comes from puzzles such as
TP2-puzzles~\cite{sun98tochi}, where two elements are inserted
concurrently just before and after an element which is being
deleted. Since deleted elements no longer separates the inserted ones,
they may be swapped.

We present a composition of two layers to ensure the ordering constraint.  We
use a set element associated with an un-mutable ordering information called
{\em position identifier (PI)}.

%We present two compositions of layers to ensure the ordering
%constraint. The first compositions use a set element associated with
%an un-mutable ordering information called {\em position identifier
%  (PI)}. The second composition use a grow-only array of elements
%associated with visibility flag. Both solutions avoid the above puzzle
%since, either the positions are totally ordered and un-mutable or the
%deleted elements still separate the others.

% When we want place an element in sequence we insert in position which is a
% integer.  But, the problem occurs when different person add and delete at same
% position.  For example, the text is ``ecd''. Alice add $a$ in places $0$ and
% Bob deletes character in place $0$.  The order of operation receiption it can
% delete 'e' or 'a'. The next insertion in $0$ could be before or after the 'a'
% or 'e'

%\subsubsection{Position identifier}

\begin{figure}[htb]
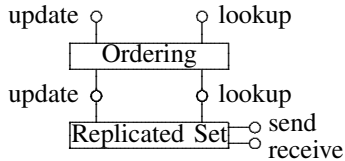

$$
\pic
\Edge (10, 8) (10, 20) (50, 8) (50, 20)
\Edge (10, 2) (10, -10) (50, 2) (50, -10)
\Vertex (10, 20) (50, 20)
\Vertex (10, -10) (50, -10)
\Rectangle (0, 0) (60, 10)
\Align[c] (Ordering) (30, 5) 
\Align[c] (update) (-10, 20) 
\Align[c] (lookup) (70, 20) 
\Translate (0, 0) (0, -30)
\Edge (10, 8) (10, 20) (50, 8) (50, 20)
\Edge(58, 2) (70, 2) (58, 8) (70, 8)
\Vertex (10, 20) (50, 20)
\Vertex(70, 2) (70, 8) 
\Rectangle(0, 0) (60, 10)
\Align[c] (Replicated Set) (30, 5)
\Align[c] (update) (-10, 20) 
\Align[c] (lookup) (70, 20) 
\Align[l] (send) (75, 10)
\Align[l] (receive) (75, 0)
%\Align[l] (To propagation) (75, -10)
%\Align[l] (algorithm) (75, -20)
\cip
$$
\caption{Text data type using sets}
\label{fig:textset}
\end{figure}

As presented in Figure~\ref{fig:textset}, we define an adaptation
ordering layer on top of a set replication layer.  The set contains
elements coupled with a position identifier (PI). For example, the
sequence 'AC' corresponds to the set $\{('A',p_a),('C',p_c)\}$. To add
'B' between 'A' and 'C', we must forge $p_b$ such that $p_a \prec p_b
\prec p_c$. The set becomes $\{('A',p_a),('C',p_c),('B',p_b)\}$.  The
``lookup'' function uses the total order between PIs to compute the
ordered sequence 'ABC'.

Position identifiers are defined in a dense space equipped with a
total ordering relation. The total order ensures that any pair of
elements appear in the same order on each replica. The space is
dense to allow insertion of an element between any two others. 

%This space
%is open, i.e. has no maximal or minimal element to allow insertion of
%elements at the beginning or the end of the text.

In the literature, such spaces already exist.
Logoot~\cite{weiss10logootundo} and
FCEdit~\cite{martin09collaborative} use integer or strings
concatenated with unique identifiers; the ordering relation is a
lexical ordering.  The Treedoc~\cite{preguica09commutative} algorithm
uses depth-first search on a binary tree as ordering. The position
identifier of Treedoc is a path in this tree with unique identifiers
to distinguish two similar paths.

The algorithms cited above generate unique identifier (unique for all
replicas). These identifiers are unique to ensure eventual
consistency.  So, when a same element is added concurrently at the
same place, it is inserted twice with two different identifiers.  For
instance, if two users aim to correct the word 'ct' into 'cat', these
algorithms add two 'a' and word becomes the 'caat'.

In our framework, the set ensures the eventual consistency. So, we can
relax the uniqueness of the position identifier. For instance, in
Logoot positions, the operation timestamp could be replaced by the
element it-self. Thus, we will obtain a different behavior than the
above algorithms since the concurrent insertion of two same element at
the same position will lead to a unique appearance.\footnote{Two 'a'
  added sequentially, for instance, in the word 'aardvark', will have
  different PIs.}  This behavior may seem more natural to users and is
the behavior (called ``accidental clean merge'') of most of the
control version system software (Git, SVN, etc.). Obviously, all
editing conflicts cannot be resolved using such approaches. However,
thank to our layered framework, one can add a semantic correction
layer such as~\cite{gu05ontology} above our own layers.

% In our example, the double 'a' become only one.  If you write
% the word 'aargh' the double 'a' is not in concurrence.  The ordering
% algorithm compute an position identifier between 'a' and 'r' this
% identifier will be deferent from 'a'.

%\begin{figure}[htb]
%\begin{lstlisting}
%interface Ordering<T>{
%	int add(T l, PI position); // Insert t with PI position
%	int del(T l, PI position); // Delete t with PI position
%	T get(int i);             // Get Element in i th
%	PI getPos(int i);         // Get i th position identifier
%	PI getNewPos(int i);  // Generate position identifier for i th element
%	Iterator <T> Iterator();  // Iterate elements
%}
%\end{lstlisting}
%\caption{Interface of ordering algorithm.}
%\end{figure}
%
%  /* compares two pi and returns -1 if pi1 < pi2, */
%	/*0 if pi1==pi2, and -1 if pi1 > pi2 */
%	int comparePi(PI pi1,PI pi2);
We define a couple object which contains a position identifier and a label.
We assume that each ordering algorithm implements the interface
described in Figure~\ref{seqInterface}.

\begin{figure}[htb]
\begin{lstlisting}
interface Ordering<L>{
	/*gets the position where the pi will be inserted in pis list.*/
	int getPos(PI pi, L label, List <Couple> pis);
	
	/*  returns an ordered list built from set of couple.*/
	List <Couple> order(Set <Couple> cs);
	
	/* generate position identifier with c1 < returned pi < c2 */
	PI generatePI(Couple c1, Couple c2);
}
\end{lstlisting}
\caption{Interface of ordering algorithm.}
\label{seqInterface}
\end{figure}

%To simplify the algotithm writing, we write three functions :
%\lstinline{int getPos(PI pi, L label, List <Couple> pis)} gets the position where the pi will be inserted in list pis.
%\lstinline{List <Couple> order(Set <Couple> cs)} returns an oredered list builded from set of couple.
%\lstinline{PI generatePI(int i,List <Couple> pis)} builds a new position $pi$ identifer like $pis[i]<pi<pis[i+1]$.  

%\begin{figure}[htb]
%\begin{lstlisting}
%/* Gets position who inserted the position pi in pis list */
%	int getPos(PI pi, L label, List <Couple> pis); 
%
% /* makes ordered list from set composed by couple */
%	List <Couple> order(Set <Couple> cs);
%
%	/* Generates new pi as pis[i] < pi < pis[i+1] */
%	PI genPI(int i,List <Couple> pis);
%\end{lstlisting}
%\end{figure}
%	/* Get position who inserted the position pi in pis list */
%	int getPos(PI pi, L label, List <C> pis); 
%  /* To order set composed by couple */
%	List <C> order(Set <C> cs);

We define the Ordering layer in two versions :  the non-incremental version in
figure~\ref{seqLayerNinc} and the incremental version in figure~\ref{seqLayer}.

The difference between two versions is the presence of the inner state.  The
non-incremental layer must order the set to have a lookup or to modify the
sequence, while the incremental version uses its inner state to avoid
re-computation.

The application or upper layer invokes the modify function of ordering layer
with operation as argument. This operation can be an add or delete operation.
%The add operation contains an element (line, characters, ...) and an integer
%position and the delete contains only an integer. 

For both layer versions, the "add" operation parameters are an element (line, characters,
...) and an integer position.  In this case, the layer gets the previous and next
element PI from the lookup list .  It generates a position identifier help with
ordering algorithm between two PIs ($generatePI$) (l.9 fig.~\ref{seqLayerNinc} and fig.~\ref{seqLayer}) and store the couple with
added element and generated position identifier in the inner set (l.15).  In case of
delete, the operation contains only the element position to remove.  The modify
function gets the element from lookup list (l.12) and forges the operation for deletion from
the inner set (l.13).

The difference between incremental and non incremental version is: for
non-incremental version, the lookup list is built from the inner set (using of
the ordering algorithm) for each call (l.6 fig.~\ref{seqLayerNinc}); while the
lookup of the incremental version returns its own up-to-date list (l.3
fig.~\ref{seqLayer}).  In incremental case, when the inner set is modified by
local or remote operation the layer is notified and update function is called.
The update function places the new element in the layer state in position given
by ordering algorithm (l.22 fig~\ref{seqLayer}) or deletes from layer state the
element which, contains the position (l.24).

%\begin{figure}[htb]
%\begin{lstlisting}
%class OrderingLayer{
%	Ordering algo;
%	
%	/* Operation sent by upper layer*/
%	void modify(Operation change){
%		Operation op;
%		if (change.type == add){
%			PI pi = algo.getNewPos(change.positon);
%			op = new Operation(add, new Couple(change.label, pi));
%		}else{
%			PI pi = algo.getPos(change.position);
%			L label = algo.get(change.position);
%			op = new Operation(add, new Couple(label, pi));
%		}			
%		innerSet.modify(op);
%	}
%
%	/* notification by inner layer*/
%	void update(Operation change){
%		C couple = change.label
%		if (change.type == add){
%			algo.add(couple.label, couple.pi);
%		}else{
%			algo.del(couple.label, couple.pi);
%		}
%	}
%}
%\end{lstlisting}
%\caption{Sequence layer}
%\label{seqLayer}
%\end{figure}

\begin{figure}[htb]
\begin{lstlisting}
class OrderingLayer{
	Ordering algo;

	void modify(SequenceOperation change){
		SetOperation op;
		List <Couple> list = lookup(); //Reordering
		if (change.type == add){		  
			int pos=change.position;
			PI pi = algo.generatePI(list.get(pos), list.get(pos+1));
			op = new SetOperation(add, new Couple(change.label, pi));
		}else{		//del operation
			Couple c = list.get(change.position);
			op = new SetOperation(del, c);
		}			
		innerSet.modify(op);
	}

	list lookup(){
	  return algo.order(innerSet.lookup);
	}
}
\end{lstlisting}
\caption{Non-Incremental Sequence layer}
\label{seqLayerNinc}
\end{figure}
%
%		String str = "";
%		for(Couple c:list){
%			str+ = c.label();
%		}
\begin{figure}[htb]
\begin{lstlisting}
class OrderingLayer{
	Ordering algo;
	List <Couple> list;

	void modify(SequenceOperation change){
		SetOperation op;
		if (change.type == add){
		  int pos=change.position;
			PI pi = algo.generatePI(list.get(pos), list.get(pos+1));
			op = new SetOperation(add, new Couple(change.label, pi));
		}else{  		// del operation
			Couple c = list.get(change.position);
			op = new SetOperation(del, c);
		}			
		innerSet.modify(op);
	}

	void update(SetOperation change){
		Couple couple = change.label
		if (change.type == add){
			int pos = getPos(couple.pi, list);
			list.add(pos, couple);
		}else{   // delete
			list.remove(couple);
		}
	}
	
	list lookup(){
	  return list;
	}
}

\end{lstlisting}
\caption{Incremental Sequence layer}
\label{seqLayer}
\end{figure}

\vspace{-1em}
\subsection{Unordered tree}
\label{sec:unordered}

In this section, we design replicated unordered trees.  The unordered tree node
contains a $Label \in \Sigma$, a father and a set of children.
The root is a special node without father and label.

\begin{figure}
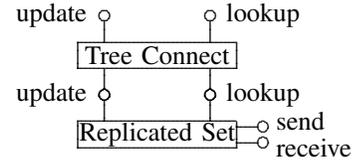

$$
\pic
\Edge (10, 8) (10, 20) (50, 8) (50, 20)
\Edge (10, 2) (10, -10) (50, 2) (50, -10)
\Vertex (10, 20) (50, 20)
\Vertex (10, -10) (50, -10)
\Rectangle (0, 0) (60, 10)
\Align[c] (Tree Connect) (30, 5) 
\Align[c] (update) (-10, 20) 
\Align[c] (lookup) (70, 20) 
\Translate (0, 0) (0, -30)
\Edge (10, 8) (10, 20) (50, 8) (50, 20)
\Edge(58, 2) (70, 2) (58, 8) (70, 8)
\Vertex (10, 20) (50, 20)
\Vertex(70, 2) (70, 8) 
\Rectangle(0, 0) (60, 10)
\Align[c] (Replicated Set) (30, 5)
\Align[c] (update) (-10, 20) 
\Align[c] (lookup) (70, 20) 
\Align[l] (send) (75, 10)
\Align[l] (receive) (75, 0)
\cip
$$
\caption{Layered tree}
\label{treelayer} 
\end{figure}

\begin{figure}[htb]
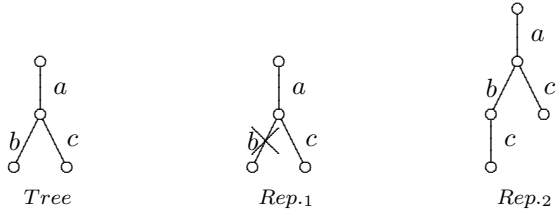

\centering
%\begin{array}{cccc}
\subfigure[$Tree$]{%\hspace{0.5em}
$\pic
\Edge(0, 0) (0, -20) (0, -20) (-10, -40) (0, -20) (10, -40)
\Vertex(0, 0) (0, -20) (10, -40) (-10, -40)
\Align[l] ($a$) (5, -10)
\Align[l] ($b$) (-12, -30)
\Align[l] ($c$) (10, -30)
\cip$%\hspace{0.5em}
}
\hspace{2cm}
\subfigure[$Rep._{1}$]{%\hspace{0.5em}
$\pic
\Edge(0, 0) (0, -20) (0, -20) (-10, -40) (0, -20) (10, -40)
\Vertex(0, 0) (0, -20) (10, -40) (-10, -40)
\Align[l] ($a$) (5, -10)
\Align[l] ($b$) (-12, -30)
\Align[l] ($c$) (10, -30)
\Line (-10, -25) (0, -35) 
\Line (0, -25) (-10, -35)
\cip$%\hspace{0.5em}
}
\hspace{2cm}
\subfigure[$Rep._{2}$]{%
$\pic
%\Edge (-10, -40) (0, -20) (10, -40)
%\Vertex(10, -40) (-10, -40)
\Edge(0, 0) (0, -20) (0, -20) (10, -40) (0, -20) (-10, -40) (-10, -40) (-10, -60)
\Vertex(0, 0) (0, -20) (10, -40) (-10, -40) (-10, -60)
\Align[l] ($a$) (5, -10)
\Align[l] ($b$) (-12, -30)
\Align[l] ($c$) (10, -30)
%\Align[l] ($b$) (5, -30)
\Align[l] ($c$) (-5, -50)
\cip$%\hspace{0.5em}
}
\caption{Concurrent operations in replicated trees}
\label{fig:conctree}
\end{figure}

As presented in Figure~\ref{treelayer}, to provide this tree, the layer uses a
set of paths. More formally, we define a path as a sequence of label: $p\in
Path, p = l_1l_2\cdots l_n, l_i \in \Sigma, \forall i \in [1..n]$.  Each path
in this set represents a node. For example, the tree drew in
figure~\ref{fig:conctree} is represented by $\{a, ab, ac\}$.
In this example, when the replica $2$ adds $c$ under $b$ the word $abc$ is
added in inner set.  When the replica $1$ removes $b$, the word $ab$ is deleted
in inner set.  In second time, both replica exchange these operations and those
states become $\{a, ac, abc\}$. This set does not represent directly a tree
because the node $b$ is not present and has one child. We call the path $abc$,
respectively the node represented by this path, an orphan path respectively an
orphan node.  In this case, there are different ways to adapt the tree from the
path set. Each way makes a different behavior.

%The layer manages the conflict between insertion
%of a node and suppression of its father.  In figure~\ref{tree}, two
%replicas share the same tree which corresponds to the inner set of
%paths $\{a, ab, ac\}$. Then, replica $1$ adds $c$ under $b$ while the
%replica $2$ concurrently removes $b$. The merged inner set state is
%$\{a, abc, ac\}$. In this state, the path $ab$ is missing, we call the
%path $abc$ an orphan path.

\begin{figure}[htb]
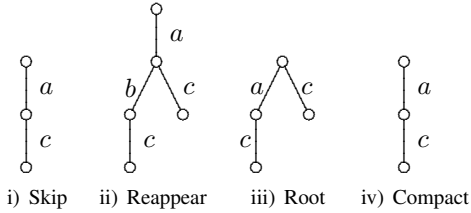

\centering
\subfigure[\mbox{Skip}]{\hspace{0.5em}
$\pic
\Vertex(0, 0) (0, -20) (0, -40)
\Edge(0, 0) (0, -20) (0, -20) (0, -40)
\Align[l] ($a$) (5, -10) 
\Align[l] ($c$) (5, -30) 
\cip$\hspace{0.5em}
}%&
\subfigure[\mbox{Reappear}]{\hspace{1.5em}
$\pic
\Edge(0, 0) (0, -20) (0, -20) (10, -40) (0, -20) (-10, -40) (-10, -40) (-10, -60)
\Vertex(0, 0) (0, -20) (10, -40) (-10, -40) (-10, -60)
\Align[l] ($a$) (5, -10)
\Align[l] ($b$) (-12, -30)
\Align[l] ($c$) (10, -30)
%\Align[l] ($b$) (5, -30)
\Align[l] ($c$) (-5, -50)
%\Edge(0, 0) (0, -20) (0, -20) (0, -40) (0, -40) (0, -60)
%\Vertex(0, 0) (0, -20) (0, -40) (0, -60)
%\Align[l] ($a$) (5, -10)
%\Align[l] ($b$) (5, -30)
%\Align[l] ($c$) (5, -50)
\cip\hspace{1em}$
\label{fig:reappearExample}
}
\subfigure[\mbox{Root}]{%\hspace{0.5em}
$\pic
\Vertex(0, 0) (10, -20) (-10, -20) (-10, -40)
\Edge (0, 0) (10, -20) (0, 0) (-10, -20) (-10, -20) (-10, -40)
%\Vertex (0, 0) (-20, -20) (20, -20)
\Align[r] ($a$) (-7, -10)
\Align[l] ($c$) (7, -10)
\Align[r] ($c$) (-12, -30)
\cip$\hspace{0.5em}
\label{fig:rootExample}
}%&
\subfigure[\mbox{Compact}]{\hspace{2em}
$\pic
\Edge(0, 0) (0, -20) (0, -20) (0, -40)
\Vertex(0, 0) (0, -20) (0, -40)
\Align[l] ($a$) (5, -10)
\Align[l] ($c$) (5, -30)
\cip%\end{array}
$\hspace{2em}
}
\caption{Different behavior for resolving conflict in trees}
\label{treebehavior}
\end{figure}

In Figure~\ref{treebehavior}, we present four different behaviours: i)
Skip behaviour does not return orphan nodes; ii) Reappear behaviour
returns the orphan node at their original path; if the node $abc$ is finally
deleted, $ab$ disappears; iii) Root behaviour places orphans under a
specific directory (root or lost-and-found); iv) Compact behaviour
moves $c$ node under node $a$, both $ac$ are merged.

More formally, we call an orphan path, a path in the inner set lookup ($LS$) that has
a prefix which is not in $LS$. We start by adding all non-orphan paths
of $LS$ to lookup of the tree ($LT$). Then, we treat the orphan paths
in $LS$ in length order (shortest first, then $\Sa$
order). Considering each orphan path $a_1a_2\ldots a_n \in LS$ with
$\forall i \in [1, n].~a_i\in \Sa$, we can apply the following {\em
  connection policies}~:

\begin{itemize}
\item[\bf skip:] {\em drops the orphan path.}
\item[\bf reappear:] {\em recreates the path leading to the orphan path.}
  We add all $a_1 \ldots a_j$ with $j \in [1, n]$.
\item[\bf root:] {\em places the orphan subtree under the root.} We add $a_j
  \ldots a_n$ to $LT$ with $j$ such that $a_1 \ldots a_{j-1} \notin
  LS$ and $\forall k \in [j, n]$, $a_1 \ldots a_{k} \in LS$.
\item[\bf compact:] {\em places the orphan subtree under its longest
    non-orphan prefix.} We add $a_1 \ldots a_ma_j \ldots a_n$ to $LT$
  with $j$ and $m$ such that $m<j$ and $a_1 \ldots a_m \in LT$ and
  $a_1 \ldots a_{m+1} \notin LS$ and $a_1 \ldots a_{j-1} \notin LS$
  and $\forall k \in [j, n]$, $a_1 \ldots a_{k} \in LS$.
\end{itemize}

% \begin{example}
%   For a lookup $LS = \{\e, a, ab, ac, abcd, abcde, abcdefg\}$, the
%   orphans path are $\{abcd, abcde, abcdefg\}$ and we obtain $LT$ equal
%   to~:
% \begin{description}
% \item[skip]     $\{\e, a, ab, ac\}$
% \item[reappear] $\{\e, a, ab, ac, abc, abcd, abcde, abcef, abcdefg\}$
% \item[root]     $\{\e, a, ab, ac, d, de, g\}$
% \item[compact]  $\{\e, a, ab, ac, abd, abde, abdeg\}$
% \end{description}
% \end{example} 

%Trees could be represented by path set, by node set with edge set (couple of
%node) or edges set with nodes set generated on the fly.  For this example we
%will build ordered tree with a ordered path set.  Like the previous example we
%use a replicated set. % or another set who the
%eventual consistency. 

\renewcommand{\thesubfigure}{}

\renewcommand{\thesubfigure}{\roman{subfigure})}

Using any of the above policies ensures that the lookup trees
presented to the client by any layered tree are eventually
consistent. Indeed, we assume that the inner set is eventually
consistent. Since the tree lookup is deterministically computed each
time the set is modified, this tree lookup is eventually
consistent. Of course, re-computing the whole tree lookup is not
efficient, and we can define incremental version of the four
policies. We present here the reappear and root incremental
policies\footnote{Due to space limitation, skip and compact policies are
not presented but are implemented in our open-source framework.}.

%%%%%policy

\subsubsection{Reappear Policy}
%The layer state is described in Figure~\ref{lStateReap}.

The reappear algorithm presented in Figure~\ref{reappearAlgo}, uses a set of
``ghosts''. When an orphan node is added in the inner set, the policy recreates
its ancestors as ghosts by browsing through the path.  When a node with children is
removed in the inner set, this node is not removed in the tree. But it is just marked as
a ghost.  Ghosts are unmarked when the node path is re-added in the set. 
All leaf nodes marked as "ghost" are recursively removed until there was nothing left.
%When a leaf node
%is a ghost, this node is removed and recursively its father, if the father is a
%ghost with no more children.  
In our example $b$ is a ghost (see Fig.~\ref{fig:reappearExample}).

The update function for the reappear algorithm is written in figure~\ref{reappearAlgo}.
The modify function converts a path of lookup to a path for inner set.  By
chance, in this policy the path is not modified. Thus, add operation is not
modified.  However, the delete operation must delete the subtree. In this case,
the algorithm looking for all children to remove from the inner set.

The update function accepts an operation which contains type of operation (add
or delete) and a path.  The path designates the new label or the label to
remove; and where to add the new node or the node to remove. The constructor
prototype of this operation is \lstinline{Operation(Optype optype, Path path)}.

%\begin{figure}[htb]
%\begin{lstlisting}
%class LayerState {
%	Layer upperL; // Upper layer
%	Node root = new Node(null, null);
%
%	Node add(Node father, L label){
%		Node n = father.childrend.containsL(label))
%			if (n == null){
%				n = new Node(father, label);
%				father.children.add(n);
%				/* Notify upper layer of addition */
%				if (upperL != null)
%					upperL.update(new Operation(add, node.getPath()));
%			}
%		return n;
%	}
%
%	void del(Node node){
%		/* Notify upper layer of deletion */
%		if (upperL != null)
%			upperL.update(new Operation(del, node.getPath()));
%		node.father.children.del(node);
%		node.father = null;
%	}
%
%	Node getNode(Path path){
%		Node node = root;
%		for(Label l:path){
%			node = node.getChild(l);
%		}
%		return node;
%	}
%}
%\end{lstlisting}
%\caption{Layer state for reappear algorithm}
%\label{lStateReap}
%\end{figure}

\begin{figure}[htb]
\begin{lstlisting} 
void Update(SetOperation change) {
	Path path = change.content;
	if (change.type == add) {	      // Adds Operation.
		Label last = path.removeLast();   // Computes the father path
		Node father = tree.getNode(Path); // Get father from path
		if (father == null) {             // If node is Orphan node
			Node node = tree.root;		   
			Path nPath = new Path();
			for (Label l: path) {          
				Node c = node.getChild(l);    
				if (c == null) {
					c = tree.add(node, l);     // reappear as ghost
					ghosts.add(c);
				}
				node = c;
			}		
			tree.add(node, last);        
		} else {                          // Not Orphan Node
			Node node = tree.add(father, last, path); 
			ghosts.remove(node);            
		}
	} else {                            // Del Operation 
		Node node = tree.getNode(path);
		if (node.children.isEmpty()) {    
			do {                            // Purge ghosts
				Node father = node.getFather(); 
				ghosts.remove(node); 	      
				tree.del(node);
				node = father;		
			} while (ghosts.contains(node) && node.children.isEmpty());
		} else {                          // Node has children
			ghosts.add(node);               // Become a ghost
		}
	}
}  
\end{lstlisting}
\caption{Update function for incremental reappear policy}
\label{reappearAlgo}
\end{figure}

%When 
      %path = new Path(change.father.Paths.get(0), change.label);
%\begin{figure}[htb]
%\begin{lstlisting}
%void modify(Operation change)
%	if (change.type == add){
%		innerSet.modify(change);
%	}else{                // Del modification
%		Node node = lState.getNode(change.path);
%		delSubtree(node, change.path);
%	}
%}
%
%/*look for and delete all path of node children*/
%void delSubtree(Node node, Path path){  
%	for(Node child:node.children){
%		delSubtree(child, new Path(path, child.label));
%	}
%	innerSet.modify(new Operation(del, path));
%}
%\end{lstlisting}
%\caption{Modify function for incremental reappear policy}
%\label{reapperModify}
%\end{figure}

\subsubsection{Root policy}
\label{sec:rootpolicy}

The root algorithm moves all orphan nodes to the root or some special
``lost-and-found'' directory.  The update function of this algorithms is
presented in figure~\ref{fig:UpdateRootAlgo}.  When two nodes with same label
are orphans, the orphans are merged and the view presents only one
node under the root. The internal state of the connecting layer is a decorated
tree. Nodes are decorated with $Paths$, the set of original paths leading to
the node. The connecting layer also uses $path2node$, a map to link original
paths to the node objects.

When a node is added, if this path is prefix of orphans paths, then all
corresponding nodes are reattached by move function.  The move function looks
for all prefixes in $Paths$ of all children of the root node and removes them.  It
adds the node to reattach and adds this prefix.  All nodes with empty $Paths$ are
deleted.

The modify function browses the tree through a path, takes the last node and
forges the operation with the $Paths$.  For example, in case of add operation,
the modify function adds each element of $Paths$ concatenated by new
label and in case of delete operation it deletes every path present is $Paths$.

%The move function generate all possibly
%path with prefix and looking for help with hash set a orphans to move.  When a
%node is already existing under father, it just add path to $Paths$. 

In our example\ref{fig:rootExample}, when $b$ is deleted and $c$ is added under $b$, the $c$ is
moved under the root. However, a node $c$ is already under the root. Two nodes $c$
fusion and $c$ contains the path $c$ and path $abc$.

%We move under $n$, all nodes which
%were formerly directly or indirectly under $n$ and the new node is added with
%its path. The map is useful to find moved nodes by path.  When an orphans node
%is added in the inner set, we add under the last visible node on its path.
%When a node with children is removed, the node is deleted and children are
%moved to father.

%The inner state for root policy is extended with path informations (see
%figure~\ref{extlstate}).
%
%\begin{figure}[htb]
%\begin{lstlisting}
%class LayerStateDecorated extends LayerState{
%	HashMap <Path, Node> p2n; //link path to node
%
%	// surcharge of add with path
%	Node add(Node father, L label, Path path){
%		Node n = add(father, label);
%		n.Paths.add(path);
%		return n;
%	}
%
%	// surcharge of del with path
%	void del(Node node, Path path){
%		node.Paths.remove(path); 
%		if (node.Paths.isEmpty()){
%			del(node);  //remove the node 
%		}
%	}
%}
%\end{lstlisting}
%\caption{Extension of layer State for incremental root policy}
%\label{extlstate}
%\end{figure}

\begin{figure}[htb]
\begin{lstlisting}
//move node identified by path from srcFather to dest
void move(Node srcFather, Node dest, List path) {     
	for (Node child: srcFather.getChildren()) {
		/* Make path with prefix and label */
		List childPath = new Path(path, child.getValue());
		/* node contains good prefix*/
		if (child.Paths.contains(childPath)) { 
			child.del(childPath);
			Node node = dest.add(child.label, childPath);
			move(child, node, childPath);
			path2node.put(childPath, node);
		}
	}
}

void Update(SetOperation change) {
	Path path = change.getContent();        
	if (change.getType() == add) { // Add
		Path fatherPath = path.clone();
		Label last = fatherPath.removeLast();
		Node father = tree.path2node.get(fatherPath);
		if (father == null){             // Orphan node
			father = tree.root;
		}
		Node node = father.add(last, path);   
		tree.path2node.put(path, node);       
		move(tree.root, node, path); // Reattach adopted 
	} else {                          // Remove
		Node node = tree.path2node.get(path);
		tree.path2node.remove(path);
		move(node, root, path);
		tree.del(node, path); //remove if paths is empty
	}
}
\end{lstlisting}
\caption{Update function for Incremental root policy}
\label{fig:UpdateRootAlgo}
\end{figure}

\subsection{Ordered Tree Data Type}
\label{sec:ordered}

\begin{figure}[hct]
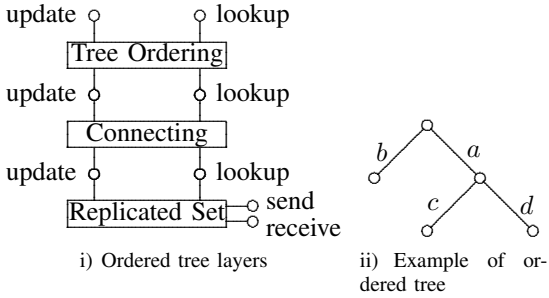

\subfigure[Ordered tree layers]{
\pic
\Edge (10, 8) (10, 20) (50, 8) (50, 20)
\Edge (10, 2) (10, -10) (50, 2) (50, -10)
\Vertex (10, 20) (50, 20)
\Vertex (10, -10) (50, -10)
\Rectangle (0, 0) (60, 10)
\Align[c] (Tree Ordering) (30, 5) 
\Align[c] (update) (-10, 20) 
\Align[c] (lookup) (70, 20) 
\Translate (0, 0) (0, -30)
\Edge (10, 8) (10, 20) (50, 8) (50, 20)
\Edge (10, 2) (10, -10) (50, 2) (50, -10)
\Vertex (10, 20) (50, 20)
\Vertex (10, -10) (50, -10)
\Rectangle (0, 0) (60, 10)
\Align[c] (Connecting) (30, 5) 
\Align[c] (update) (-10, 20) 
\Align[c] (lookup) (70, 20) 
\Translate (0, 0) (0, -60)
\Edge (10, 8) (10, 20) (50, 8) (50, 20)
\Edge(58, 2) (70, 2) (58, 8) (70, 8)
\Vertex (10, 20) (50, 20)
\Vertex(70, 2) (70, 8) 
\Rectangle(0, 0) (60, 10)
\Align[c] (Replicated Set) (30, 5)
\Align[c] (update) (-10, 20) 
\Align[c] (lookup) (70, 20) 
\Align[l] (send) (75, 10)
\Align[l] (receive) (75, 0)
%\Align[l] (To propagation) (75, -10)
%\Align[l] (algorithm) (75, -20)
\cip
%\caption{Ordered tree layers}
\label{fig:ordered}
}
\subfigure[Example of ordered tree]{
\pic 
\Edge (0, 0) (20, -20) (0, 0) (-20, -20) (20, -20) (0, -40) (20, -20) (40, -40)
\Vertex (0, 0) (20, -20) (-20, -20) (0, -40) (40, -40)
\Align[l] ($a$) (15, -10)
\Align[r] ($b$) (-15, -10)
\Align[l] ($c$) (0, -30)
\Align[l] ($d$) (35, -30)
\cip
%\caption{Example of ordered tree}
\label{fig:otree}
}
\caption{Ordered tree}
\end{figure}

In this section, we design ordered tree. As presented in Figure~\ref{fig:ordered},
we directly use the unordered tree data structure and we add an ordering layer.
To order the children of a node we use {\em Position Identifier} (introduced in
Section~\ref{order}).  We mark all labels with a position identifier.
Therefore, the nodes become totally ordered. The set of paths, managed by the
replication layer, is represented by $p = (l_1, p_1) \cdots (l_n, p_n)$ with
$l_i\in \Sigma$ a label and $p_i$ a position identifier.
However, the modify interface of the tree ordering layer must be
independent of the chosen ordering algorithm. The ordering layer
interface receives operation based on a path defined on integer position
without label (ex : 2.4.5.1).  Each integer position corresponds to a
children number in the ordered tree.  For example, consider the tree
on the Figure~\ref{fig:otree}.  The inner replicated set contains
$\{a_{p_a}, b_{p_b}, a_{p_a}c_{p_c}, a_{p_a}d_{p_d}\}$ with $p_b \prec
p_a$ and $p_c \prec p_d$. The ordered path leading to $c$ is $2.1$.

In fact, in a similar way as an unordered tree, the layer state contains nodes,
but, each node, contains additionally the position identifier and each child is
ordered by chosen ordering algorithm.

The modify function converts an integer position path $j_1...j_n$, $j_i \in
\mathbb{N}$ into a path containing couples of label and position identifier.
It browses through the tree and pushes the couple of label and position
identifier for each node, until the last but one. If the operation is an add,
the last position identifier $pi_n$ is generated by ordering algorithm.  The
generated position identified by $pi_n$ where $pi_{j_n} \prec pi_n \prec pi_{{j_n}+1}$
if $j_n$ is the last position of path and $p_{j_n}$ is position identifier in
position $j_n$.  This holds as the last position of the path is the new node.  In
case of delete operation, the modify function converts all of path.

The update function receives a path with label and positions identifier from the
inner set.  It browses through the tree until the last node but one of the
path. The algorithm can use a Hashmap or dichotomy algorithm to find a node in
the children ordered list.  In case of add operation, the update function adds
the new node in good place defined by ordering relation.  In case of delete,
the update function deletes the node.

\subsection{Extension to schema}
\begin{figure}[h]
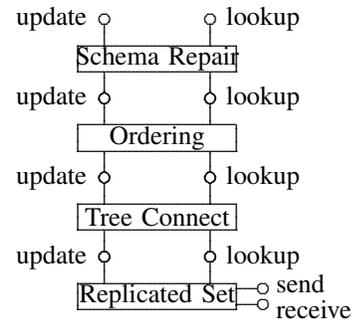

\centering
$
\pic
\Edge (10, 8) (10, 20) (50, 8) (50, 20)
\Edge (10, 2) (10, -10) (50, 2) (50, -10)
\Vertex (10, 20) (50, 20)
\Vertex (10, -10) (50, -10)
\Rectangle (0, 0) (60, 10)
\Align[c] (Schema Repair) (30, 5) 
\Align[c] (update) (-10, 20) 
\Align[c] (lookup) (70, 20) 
\Translate (0, 0) (0, -30)
\Edge (10, 8) (10, 20) (50, 8) (50, 20)
\Edge (10, 2) (10, -10) (50, 2) (50, -10)
\Vertex (10, 20) (50, 20)
\Vertex (10, -10) (50, -10)
\Rectangle (0, 0) (60, 10)
\Align[c] (Ordering) (30, 5) 
\Align[c] (update) (-10, 20) 
\Align[c] (lookup) (70, 20) 
\Translate (0, 0) (0, -60)
\Edge (10, 8) (10, 20) (50, 8) (50, 20)
\Edge (10, 2) (10, -10) (50, 2) (50, -10)
\Vertex (10, 20) (50, 20)
\Vertex (10, -10) (50, -10)
\Rectangle (0, 0) (60, 10)
\Align[c] (Tree Connect) (30, 5) 
\Align[c] (update) (-10, 20) 
\Align[c] (lookup) (70, 20) 
\Translate (0, 0) (0, -90)
\Edge (10, 8) (10, 20) (50, 8) (50, 20)
\Edge(58, 2) (70, 2) (58, 8) (70, 8)
\Vertex (10, 20) (50, 20)
\Vertex(70, 2) (70, 8) 
\Rectangle(0, 0) (60, 10)
\Align[c] (Replicated Set) (30, 5)
\Align[c] (update) (-10, 20) 
\Align[c] (lookup) (70, 20) 
\Align[l] (send) (75, 10)
\Align[l] (receive) (75, 0)
%\Align[l] (To propagation) (75, -10)
%\Align[l] (algorithm) (75, -20)
\cip
$
\caption{Tree with schema}
\label{fig:dtd}
\end{figure}

In this section, we consider ordered trees with schema (such as XSD or
DTD for XML documents). Concurrent modifications can produce a tree
which does not respect the schema.  For example, consider a schema
which accepts zero to one title element. If two users add concurrently
a title, they will create two title nodes in the internal tree data
type. To fix it, we add a new layer called schema repair.
In this layer (see Fig.~\ref{fig:dtd}), lookup interface calls a repair algorithm (such
as~\cite{StCh06}) to return a valid tree. The ``modify'' must ensure
that each operation generated on lookup view is valid on internal data
structure~\cite{martin10fixing}.

For example, in an agenda, we assume that under a participants node, there is
one or more person. If there are two persons and two replica delete one distinct,
then each replica has generated an operation compatible with the schema.
However, at the end, no person is present.  The repairing algorithm has two choices:
add a person or delete participants markup. 
However, if the schema needs
participants under event node, then the algorithm chooses to add a person.  In
this case, each replica will repair by adding a person node. This addition will not
be passed to the inner data type. In our model the lookup or update does not modify the inner state.
When a node is added under the virtual person
node like a name, the modify function creates the missing node before to add
name, because the participant is not present in the inner state.  An addition under
non-present node implies a fix in tree layer. If the chosen policy is different
from reappear the result is not compliant with the schema and the tree will be
fixed again.

%*** TODO : plus imag\'e ***
%
%
%For example, the repair algorithm adds a node $r$. By definition, the
%lookup has no effect on inner data type. Therefore, this
%modification is not yet passed on inner data type. When the update
%is invoked to add $s$ under $r$, it must add $r$ before $s$ to respect
%the pre-condition of the inner data type.  In another case, the layer
%called tree fixing will wrongly hide or move $s$.

%Tu vas me dire que le fait du non respect de la précondition de la couche
%inférieur  n'est pas si grave. Sauf que sa couche (skip, reappear, etc..) va
%essayer de la réparer ce qui va poser problème.

% For the
%state based the repair algorithm~\cite{StCh06} give a fixed tree and when the user modify it
%a new version which broadcaster respects the schema.
%For operational based we can generalize the functions {\em sendop} and {\em
%fix} defined by Martin \& all in \cite{martin10fixing} for special tree.  
%An
%special operation is sent when a site respect the schema and no operation is
%queued. This operation launch the fix algorithm on all sites. 
%The nodes added
%by repair algorithm is sent when this node is needed for modification, like an
%adding a children.

%$$
%\pic
%\Edge 
%$$
\paragraph*{Optimization with DTD schema}
The particularity of DTD schema is a poor language. An add or remove of a
node can invalidate only a part of the tree. It's possible to use a
sub-quadratic algorithm~\cite{Wu1995346} to approximate regular expression
matching on children to fix the tree.  All added edges by this algorithm could
be added with a template of recursive valid children.

\subsection{Directed acyclic graph}

\begin{figure}[h]
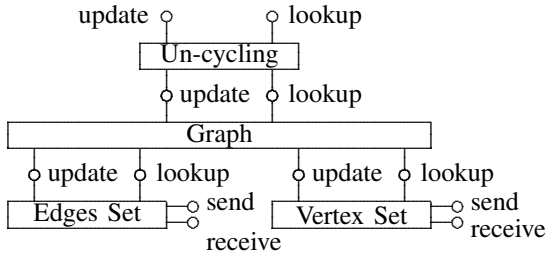

\centering
$
\pic
\Edge (10, 8) (10, 20) (50, 8) (50, 20)
\Edge (10, 2) (10, -10) (50, 2) (50, -10)
\Vertex (10, 20) (50, 20)
\Vertex (10, -10) (50, -10)
\Rectangle (0, 0) (60, 10)
\Align[c] (Un-cycling) (30, 5) 
\Align[c] (update) (-10, 20) 
\Align[c] (lookup) (70, 20) 
\Translate (0, 0) (0, -30)
\Edge (10, 8) (10, 20) (50, 8) (50, 20)
\Edge (-40, 2) (-40, -10) (0, 2) (0, -10)
\Edge (100, 2) (100, -10) (60, 2) (60, -10)
\Vertex (10, 20) (50, 20)
\Vertex (-40, -10) (0, -10)
\Vertex (100, -10) (60, -10)
\Rectangle (-50, 0) (110, 10)
\Align[c] (Graph) (30, 5) 
\Align[l] (update) (15, 20) 
\Align[c] (lookup) (70, 20) 
%\Align[c] (update) (-10, 20) 
%\Align[c] (lookup) (70, 20) 
%
\Translate (0, 0) (-50, -60)
%\Edge (10, 8) (10, 20) (50, 8) (50, 20)
%\Edge (10, 2) (10, -10) (50, 2) (50, -10)
%\Vertex (10, 20) (50, 20)
%\Vertex (10, -10) (50, -10)
%\Rectangle (0, 0) (60, 10)
%\Align[c] (Ordering) (30, 5) 
%\Align[l] (update) (15, 20) 
%\Align[c] (lookup) (70, 20) 
%
%\Translate (0, 0) (-50, -90)
\Edge (10, 8) (10, 20) (50, 8) (50, 20)
\Edge(58, 2) (70, 2) (58, 8) (70, 8)
\Vertex (10, 20) (50, 20)
\Vertex(70, 2) (70, 8) 
\Rectangle(0, 0) (60, 10)
\Align[c] (Edges Set) (30, 5)
\Align[l] (update) (15, 20) 
\Align[c] (lookup) (70, 20) 
\Align[l] (send) (75, 10)
\Align[l] (receive) (75, -5)
\Translate (0, 0) (50, -60)
\Edge (10, 8) (10, 20) (50, 8) (50, 20)
\Edge(58, 2) (70, 2) (58, 8) (70, 8)
\Vertex (10, 20) (50, 20)
\Vertex(70, 2) (70, 8) 
\Rectangle(0, 0) (60, 10)
\Align[c] (Vertex Set) (30, 5)
\Align[l] (update) (15, 20) 
\Align[c] (lookup) (70, 20) 
\Align[l] (send) (75, 10)
\Align[l] (receive) (75, 0)
%\Align[l] (To propagation) (75, -10)
%\Align[l] (algorithm) (75, -20)
\cip
$
\caption{Directed acyclic graph}
\label{fig:graph}
\end{figure}

This kind of data type can be used for task dependence representation,
such as Gantt or Pert diagram. In this example, we use two replicated
sets: a set of nodes and a set of edges. The nodes represent the
tasks, and the edges represent the dependency between the tasks. Two
concurrent dependency additions conflict when they introduce a
cycle in the graph. An un-cycling layer resolves such conflict by
traversing the graph using a breath-first search (see Fig.~\ref{fig:graph}).

% Two sets are needed because a task can be independent.

% todo example of uses case.

\vspace{-1em}
\section{Experimental Evaluation}
\label{sec:expe}

To evaluate the performances of our approach, we have implemented it
in the framework {\sf ReplicationBenchmark} developed in Java,
available on the GitHub
platform~\footnote{\url{http://github.com/score-team/replication-benchmarker}}
under the terms of the GPL license. In this framework, we have
implemented different set layers, different ordering algorithms, the
connecting layer with the four policies described
Section~\ref{sec:unordered} and the tree ordering layer described
Section~\ref{sec:ordered}.
 
The framework follow our layer structure. For instance, creating a
ordered tree based on a reappear policy and a counter replicated set
is done by the following Java expression: {\sf new
  PositionIdentifierTree(new WordTree(new ReappearPolicy(), new
  CounterSet()))}.  The framework provides base classes for common
elements, such as a version vector, set, tree and ordered tree
operations. 

The framework provides a simulator that generates a trace of operations
randomly, according to provided parameters such as trace length,
percentage of adding, removing, number of replica, communication
delay, etc.  It also provides a controlled simulation environment that
replays a trace of operations and measures the performance of the
replicated algorithms. The simulation ensures that each replica
receives operations in the order as defined in the logs. The
framework lets replicas of every algorithm generate operations in its
own formats for the given trace operations provided from the simulated
logs. The trace obtained to run our experiment has 30000 operations 
with 88\% of insertions and four replicas. The trace is available on the
 web~\footnote{\url{http://www.loria.fr/~mahmedna/trace}}.

We denote a \textit{local operation} an operation appearing in the
trace. Such operation will be given to the modify interface. For
ordered tree, operations are insertion of an element or deletion of a
sub-tree. A local operation is divided into one to several
\textit{remote operation} that the simulation sends to remote
replicas. A replica, therefore, executes remote operation. We measure
the net execution time of local and remote operations for each
algorithm. The framework uses {\sf java.lang.System.nanoTime()} for
the measurement of execution time of each local operation and each
remote operation.
 
To obtain a correct result, we ran each algorithm on traces three
times on the same JVM execution. We also measure the size memory
occupied by each algorithm. We serialize each document replica by
using Java serialization after each hundred operations generated, and
measure the size of the serialized object.

All executions are run on the same JVM, on a dual-processor machine
with Intel{\small (R)} Xeon{\small (R)} 5160 dual-core processor (4Mb
Cache, 3.00 GHz, 1333 MHz FSB), that has installed GNU/Linux
2.6.9-5. During the experiment, only one core was used for
measurement. All graphics are smoothed by bezier curves.

Before the representation out result of the experiment, we briefly describe some
representative algorithms that exist and which we will compare our
approach.

\subsection{TreeOpt and OTTree} 

TreeOPT (tree OPerational Transformation) \cite{ignat03treeopt} is a general
algorithm designed for hierarchical documents and semi-structured documents.
Each node contains an instance of an operation transformation
algorithm \cite{ellis89concurrency,ResselCSCW96,SunCSCW98}.  The algorithm
applies the operational transformation mechanism recursively over the different
document levels.  In our experimentation, we have used this algorithm with
SOCT2 \cite{SuleimanGROUP97} algorithm and TTF (Tombstone Transformation
Functions) approach \cite{oster06tombstone}. For little optimization, we
save only insertion operation in log of SOCT2.

The OTTree, an  unpublished algorithm, uses only one instance of SOCT2 for
entire the tree (not on each node) and TTF on each children list. The operation
of TTF and its integration function were modified to include the path
information.

\subsection{FCEdit} FCEdit \cite{martin09collaborative} is a CRDT designed for
collaborative editing of semi-structured documents. It associates to each
element a unique identifier.  FCEdit maps $identifier \rightarrow node$. So it
uses just an hash table to find an element in the tree. Each child is
ordered by a position identifier. Unlike OTtree, FCEdit does not need to store an
element in tombstone. The elements are really deleted from tree making it more
efficient in memory.

In the following, we present behaviors of each ordered tree algorithms
executed on simulated traces with the different policies described in
Section~\ref{sec:examples}.

\subsection{Execution times}

In \cite{JayTOCHI07}, studies have shown that
users can comfortably observe modifications on their application if
the local and remote response time do not exceed $50\,ms$. In
this section, we address an experimental evaluation of algorithms based
on our layer structure, compared to existing ones to verify if this
design is suitable for real time collaborative applications.
 
\subsubsection{Skip policy}
\paragraph{Local operations}

\begin{figure}[ht]
{\centering
\includegraphics[width = \linewidth]{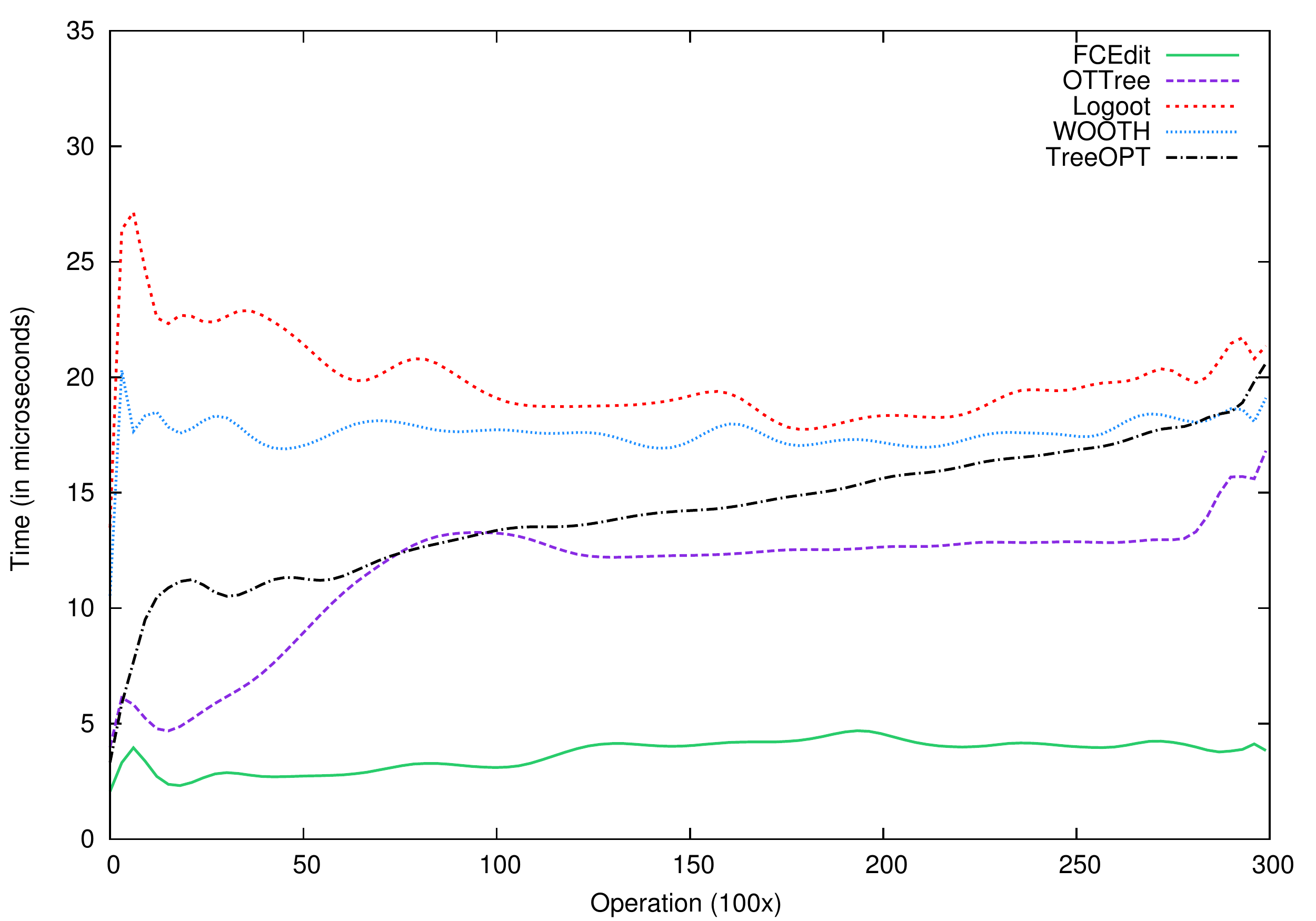}}
\caption{Execution time for algorithms with Skip -local-}
\label{fig:SkipUnique_gen}
\end{figure}

The average execution time of Local operations are presented in figure
\ref{fig:SkipUnique_gen}.

The performances of the algorithms based on the layer structure (Logoot and
WOOTH) are the less efficient compare to the algorithms that exist (OTTree and
FCEdit), but it remains stable throughout the experiment. They do not exceed
$30\mu s$, and thus $50\,ms$, what makes them acceptable for the users. The
performances of OTTree and TreeOPT based on SOCT2 algorithm degrade in the
beginning of experiment, since the rate of insertion is greater than the
deletion, the tree becomes quickly large.  TreeOPT makes an operation by each
element of the path contrary to OTTree. This explains that the difference of
both algorithms depends of tree depth.  After the 100 000 operations, the majority of  
algorithms become stable. FCEdit is the best algorithm since each node is
identified by an unique identifier, using a hash table to link identifiers and
node, they obtain a result with a complexity around $O(1+n/k)$ in the average
case. Such a ''trick'' is only possible since FCEdit uses a unique identifiers.

The global performance behaviors of Logoot and WOOTH are quite similar, even if
they are very different algorithms. This proves that the layer structure cost
in performance, but this remains stable and does not exceed $50\,ms$.

\begin{figure}[ht]
{\centering
\includegraphics[width = \linewidth]{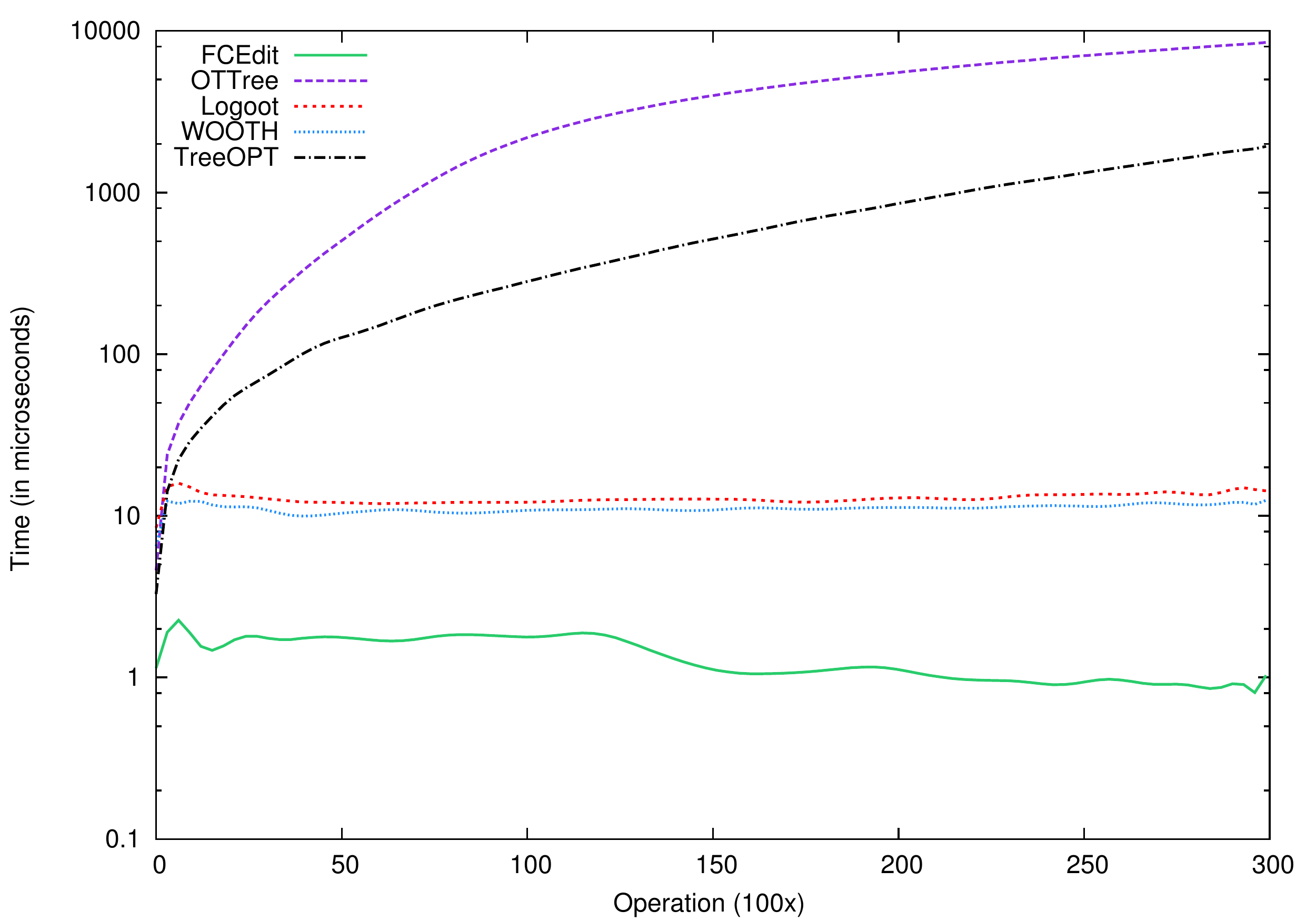}}
\caption{Execution time for algorithms with Skip -remote-}
\label{fig:SkipUnique_usr}
\end{figure}

\paragraph{Remote operations}
In Figure \ref{fig:SkipUnique_usr} we present an execution time behaviours of
algorithms using a skip policy for the remote operations on logarithmic scale. 

To simulate a real experiment, the garbage collection mechanism of SOCT2 is
disabled. Indeed, when users may disconnect, a garbage collection mechanism of
SOCT2 cannot purge the history. The performances of OTTree and TreeOPT degrade
over time since SOCT2 algorithm can not purge the history. Thus, the whole of
operations received are stored in the history and it takes time to separate
concurrent operations and transforms them that makes the algorithm the least
efficient.  Indeed, even if some garbage collection mechanisms exist, 
we consider that they can not be used in a general context
where the number of replicas is unknown and fluctuating. As locally, the
behaviors of Logoot and WOOTH algorithms remains stable, although these
algorithms are based on layer structure, they outperform OTTree and TreeOPT
with $10\mu s$ compare to $10\,ms$. 
The performance of FCEdit remains good and stable
during all experiments, with just $3\mu s$ it represents the best algorithm in
our experiment.

\subsubsection{Compare policies}
In what follows, we will present the behaviors of Logoot algorithm with
different policies that exist and also WOOTH with reappear policy. For
 ordered tree based on WOOTH algorithm,  a root and compact policies are not
permitted. Because, we cannot merge different nodes that depends by
their previous and next element with another located in different origin. 
\paragraph{Local operations}

\begin{figure}[ht]
{\centering
\includegraphics[width=\linewidth]{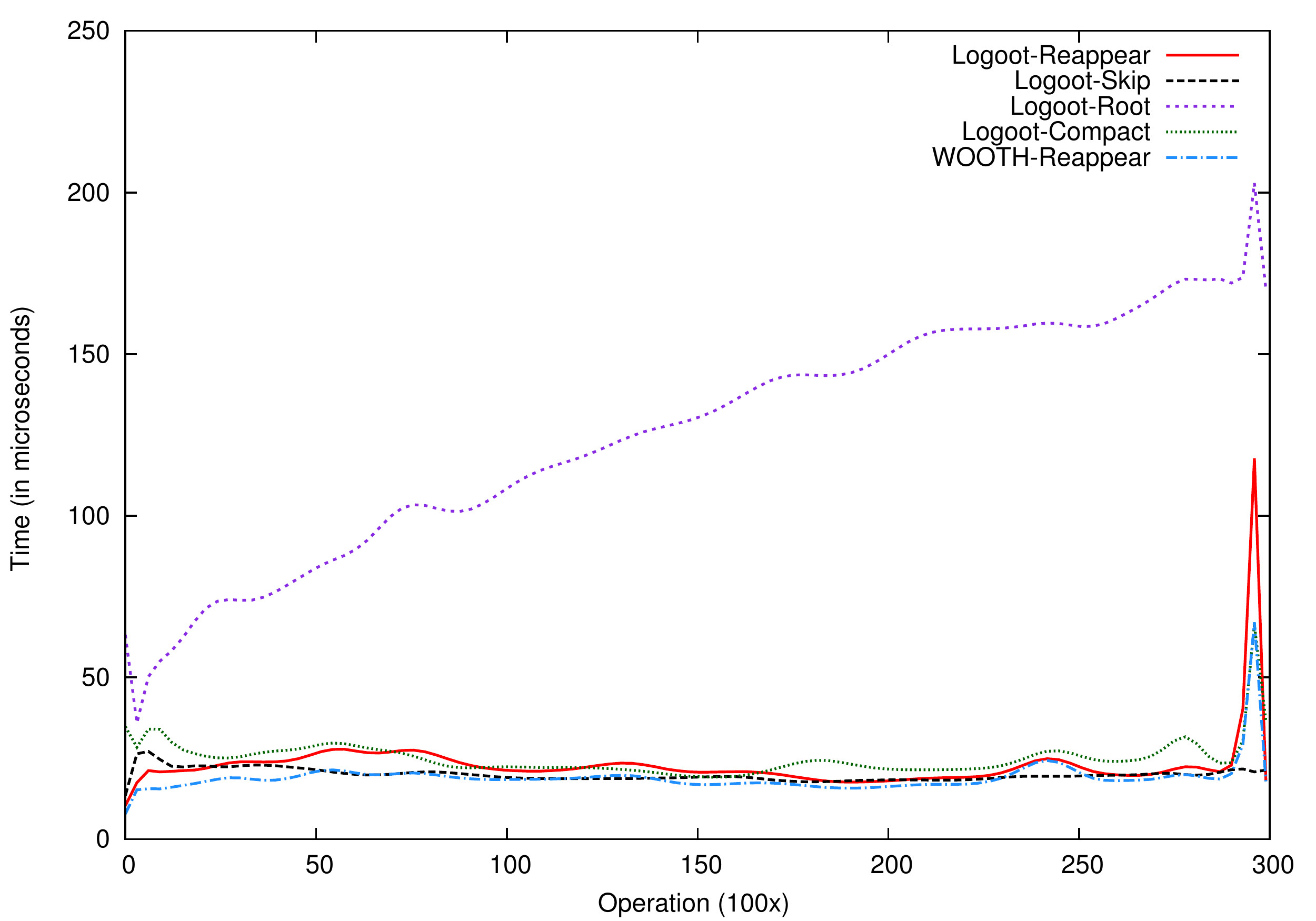}}
\caption{Execution time for algorithms with policies -local-}
\label{fig:policies_gen}
\end{figure}

In Figure \ref{fig:policies_gen} the global performance behaviors are the same
 excepted for root policy. In both policies, the algorithm must move all
 subtree deleted. In case of root policy, it moves under the root while for
compact policy it moves under the last father on the tree. In the case where
the node located in the origin path has a same label as the node in the new
path, the two nodes are merged. Since, number of nodes located under the root in
root policy are greater than the number of children under a node in compact policy,
the time lost to find the nodes with the same label in root policy takes more 
time than for compact policy. Indeed, all nodes deleted in the tree are located under
the root whereas in compact policy, a node contains his children and the nodes removed
from their child.

% For every experiment, peaks of low performance common to all algorithms exist,
% for instance in 57 00 and 242 00 operation. Such peaks are all due to
%operation
% which requiert cross the tree.

\begin{figure}[ht]
{\centering
\includegraphics[width=\linewidth]{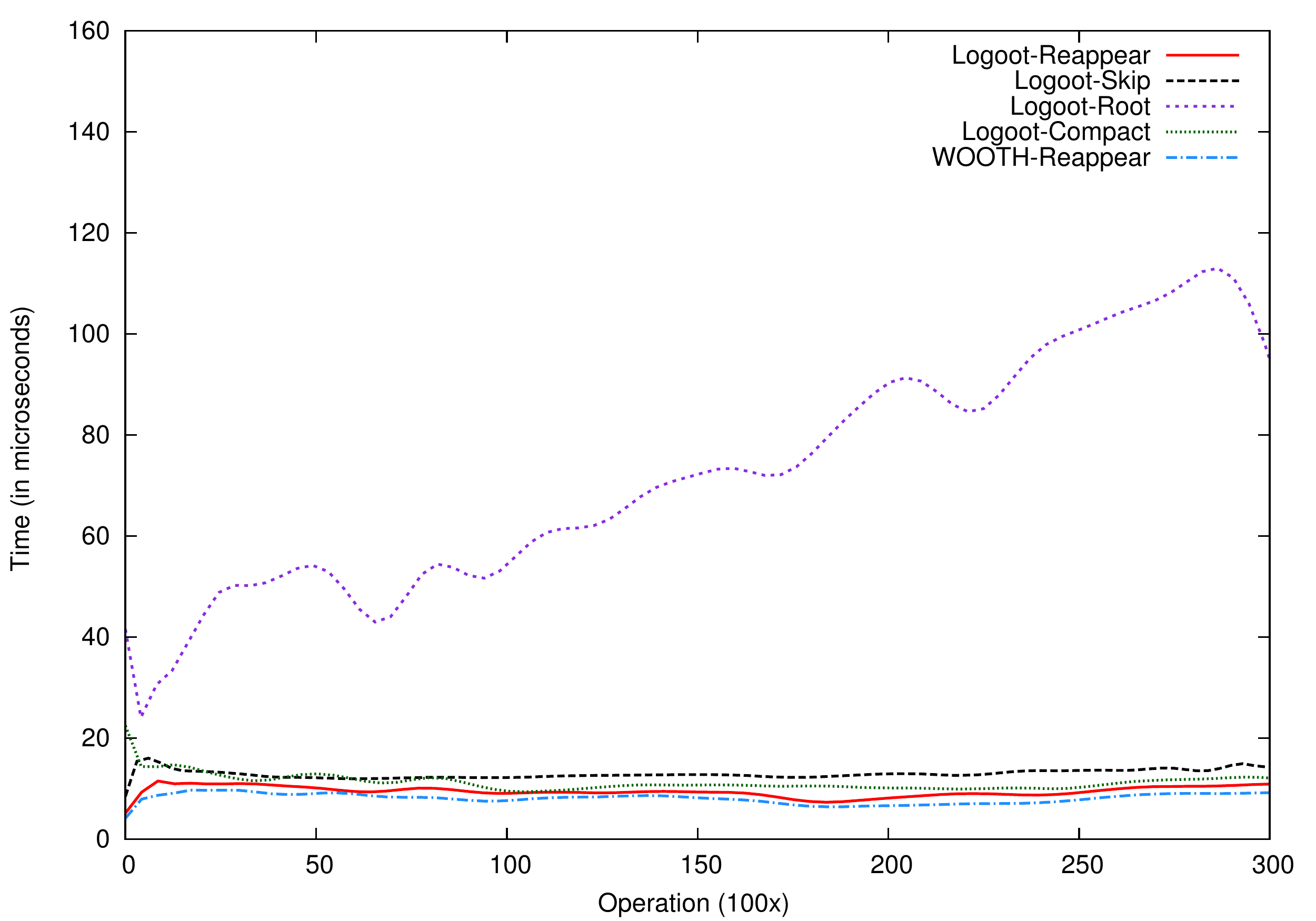}}
\caption{Execution time for algorithms with policies -remote-}
\label{fig:policies_usr}
\end{figure}

\paragraph{Remote operations}
The behavior of the different algorithm for remote operation presented in 
\ref{fig:policies_usr} is a slightly different compared to figure
\ref{fig:policies_gen} since the behaviors are more chaotic for the root policy.

The behavior of Logoot with skip policy is the most stable. The average time of
execution remains around $10\mu s$. As previously, the root behavior is the
least efficient and the most chaotic. It improves when a replica deletes a path
from the tree, as in operation number 6000 or 23000. In both algorithms Logoot
and WOOTH with Reappear policy and also Logoot with compact policy have a
chaotic behaviors although it remains stable globally.

Finally, although some algorithms are less efficient than other, the execution
time never exceeds $1ms$ (far below 50ms). And almost every algorithm has a
very stable behaviour below $30\mu s$. The Algorithms based on layer structure
are acceptable and suitable for real-time collaboration. Moreover, they
outperform some representative operational transformation as OTTree.

\subsection{Memory occupation}
\begin{figure}[ht]
{\centering
\includegraphics[width = \linewidth]{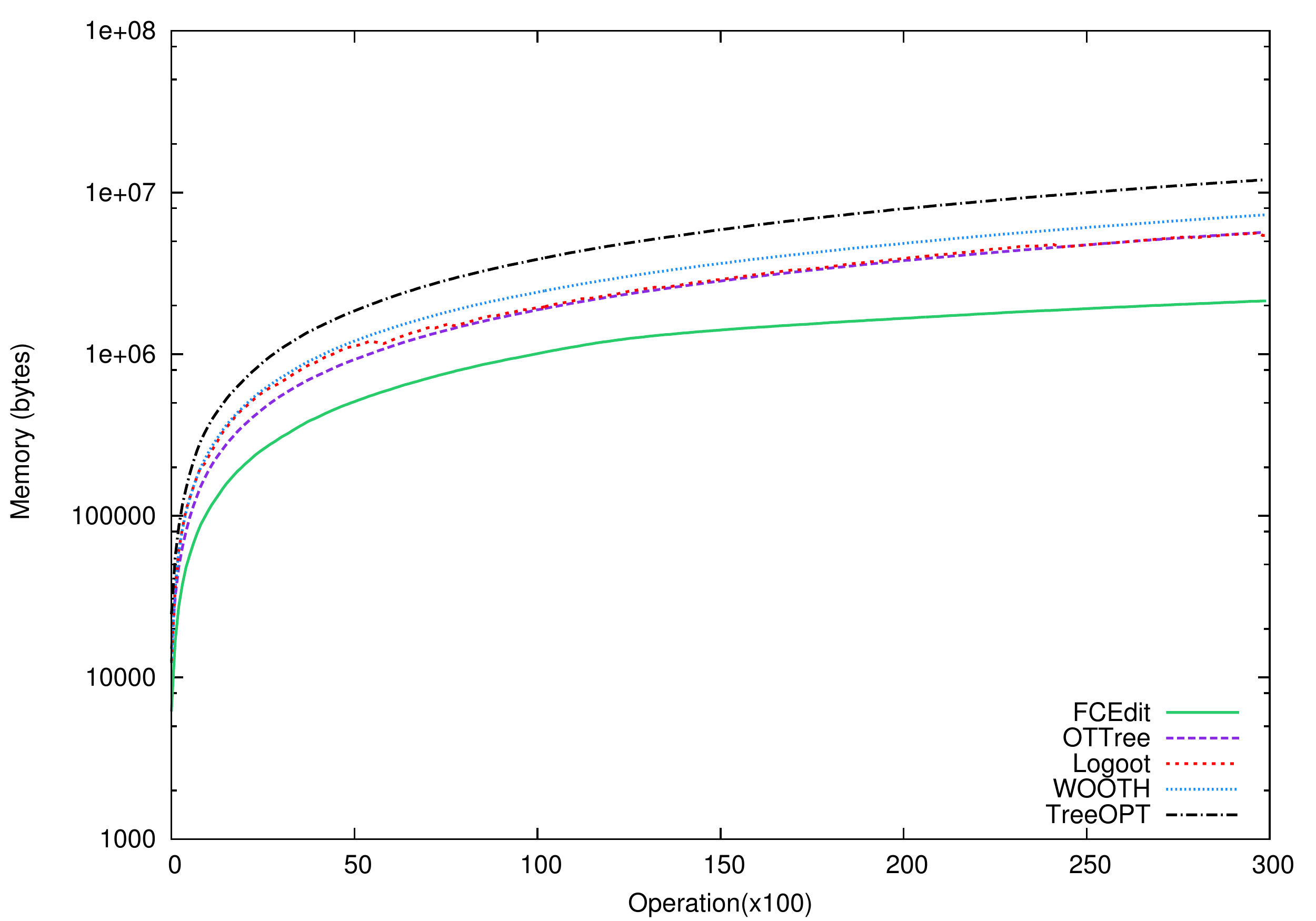}}
\caption{Memory occupation for algorithms with Skip}
\label{fig:SkipUnique_mem}
\end{figure}

Size of memory occupied by each studied algorithm may increase over time due to
history, tombstones or growing identifiers. We
present in the following, the algorithms behavior regarding memory usage in case of
skip policy on logarithmic scale illustrated in figure \ref{fig:SkipUnique_mem}.
%  and others policies in \ref{fig:policies_mem}.

A tree based on WOOTH algorithm occupies more memory compared to other tree
algorithms, since in WOOTH an identifier is never deleted but just stored in
tombstone and marked as invisible to users. OTTree and tree based on Logoot
algorithm have almost the same behavior. The memory size occupied by Logoot
depends of the size of identifiers Logoot, whereas OTTree depends of number
operation generated. Indeed, SOCT2 used in OTTree stores all operations in
history, in addition, the garbage collector was quenched, moreover a deleted
node is never removed. TreeOPT consumes more memory than OTTree because each
node has a SOCT2 instance with a log.
FCEdit remains the best algorithm regarding the memory space requirement since
the identifiers are less cost than Logoot and the nodes removed are really
deleted contrary to WOOTH and OTTree.

% 
% \begin{figure}[ht]
% {\centering
% \includegraphics[width=\linewidth]{figs/policies_mem.pdf}}
% \caption{Memory occupation for algorithms with other policies}
% \label{fig:policies_mem}
% \end{figure}

\section{Related work}

Some collaborative system, such as version control system (Git, SVN,
etc.), or distributed file systems~\cite{richard1992ficus} relies on
human merging phases for some conflict cases, while some conflicts are
resolved automatically.  For instance, SVN creates a ''tree conflict''
when a file is created in a concurrently deleted directory. On the
other hand, Git behavior is similar to ``reappear policy'' (see
Section~\ref{sec:unordered}) since it recreates silently the directory.
However, human conflict resolving does not scale to massive
collaboration use cases, and complex data types conflicts may be
difficult to represent and resolve. For instance, Git is unable to
merge correctly XML files. Our approach computes automatically a best
effort merge, and can be combined to awareness
mechanisms~\cite{dourish92awarness} to allow users to be conscious of
concurrent modifications.

There exists many systems which satisfy the {\em eventual consistency}
properties. Industrial systems, such as No-SQL data-stores (Amazon S3,
CouchDB, Cassandra, etc.), relies on eventual consistency, but only
manage key-value data types. Bayou~\cite{terry95bayou} and
Icecube~\cite{kermarrec01icecube} systems use constraints resolution
mechanisms to resolve the conflicts. So, they can ensure generic data
types constraints. But, these approaches do not scale well since they
require a central or primary server and, as in version control
systems, the system is not stable as soon as the update are delivered,
since their merge procedures produce new operations.

Replicated data types are well-known in the literacy. For instance,
there exists sets~\cite{shapiro11conflictfree},
%dictionaries~\cite{wuu84efficient}, `
sequences~\cite{oster06tombstone,
  weiss09logoot}, trees~\cite{martin10scalable}, file
systems~\cite{richard1992ficus}, etc.  In Operational Transformation
(OT)~\cite{ellis89concurrency}, replicas transform received operations
against concurrent ones. The OT approach has been successfully applied
on several general public collaborative editing software, including
Google Docs.  Conflict-free Replicated Data Types
(CRDT)~\cite{shapiro11conflictfree} aims to design replicated
data-types that integrate remote modifications without
transformation. The goal of our approach is encapsulate any eventually
consistent approach (OT or CRDT) in a replication layer and to design
adaptation layer provide to satisfy non-trivial constraints. For
instance, in our implementation (see Section~\ref{sec:expe}), we have
implemented and tested trees layers on top of both
different CRDT sets and OT sets.

%Montrer comment faire un arbre ?
%Faire un dessin 
%\begin{figure}
%$ \pic0
%\Edge (
%\Rectangle 
%\section{Conclusion}
%
%This system needs some properties on data type operations or data type state to
%ensure the eventually consistency like commutativity. It is hard to design a
%rich data type directly in this approach and the eventual consistency proof is
%as complex as data structure.  In another hand each modelling provides
%different behaviour.  The aim of this paper is to show a method for engineer to
%forge from these models a desired data type with custom conflict resolution
%while ensure the eventual consistency. 

%In this paper we have presented how to design a complex data structure from
%simple sets. Our approach composes one or several existing replicated elements
%with adaptation layer to make a rich replicated data type. 

\section{Conclusion} 

In this paper, we have presented a layered approach to design
eventually consistent data types. Our approach composes one or several
existing replicated data types which ensure eventual consistency,
and adaptation layers to obtain a new eventually consistent data
type. Each layer or replicated data type can be freely substituted by
one providing the same interface.

We have demonstrated that our approach is implementable and obtains
acceptable performances, even if these performance are sometimes
slightly worse than some specific algorithms. Our experiments and
implementation are public available and re-playable. Compared to
existing solutions, the composition design can fit precisely the
distributed application engineer wishes in terms of behavior and
scalability.

In the future works, we will run experiments on a real data like git
software histories and we will formally establish the equivalence
proof between incremental and non-incremental algorithms.

%The aim of this paper is to show a method for engineers to
%forge from these models a desired data type with custom conflict resolution
%while ensure the eventual consistency. 
\section*{Acknowledgement}
This work is partially supported by the ANR national research
grants STREAMS (ANR-10-SEGI-010) and ConcoRDanT (ANR-10-BLAN 0208).

\bibliographystyle{abbrv}
\bibliography{theBib,bib,ref}

\begin{thebibliography}{10}

\bibitem{dourish92awarness}
P.~Dourish and V.~Bellotti.
\newblock Awareness and coordination in shared workspaces.
\newblock In {\em Proceedings of the 1992 ACM conference on Computer-supported
  cooperative work}, CSCW '92, pages 107--114, New York, NY, USA, 1992. ACM.

\bibitem{ellis89concurrency}
C.~A. Ellis and S.~J. Gibbs.
\newblock Concurrency control in groupware systems.
\newblock In J.~Clifford, B.~G. Lindsay, and D.~Maier, editors, {\em SIGMOD
  Conference}, pages 399--407. ACM Press, 1989.

\bibitem{gilbert02brewer}
S.~Gilbert and N.~Lynch.
\newblock Brewer's conjecture and the feasibility of consistent, available,
  partition-tolerant web services.
\newblock {\em SIGACT News}, 33:51--59, June 2002.

\bibitem{gu05ontology}
N.~Gu, J.~Xu, X.~Wu, J.~Yang, and W.~Ye.
\newblock Ontology based semantic conflicts resolution in collaborative editing
  of design documents.
\newblock {\em Advanced Engineering Informatics}, 19(2):103 -- 111, 2005.

\bibitem{richard1992ficus}
R.~G. Guy, J.~S. Heidemann, and T.~W. Page, Jr.
\newblock The ficus replicated file system.
\newblock {\em SIGOPS Oper. Syst. Rev.}, 26(2):26--, April 1992.

\bibitem{ignat03treeopt}
C.-L. Ignat and M.~C. Norrie.
\newblock Customizable collaborative editor relying on treeopt algorithm.
\newblock In {\em Proceedings of the eighth conference on European Conference
  on Computer Supported Cooperative Work}, ECSCW'03, pages 315--334, Norwell,
  MA, USA, 2003. Kluwer Academic Publishers.

\bibitem{imine03proving}
A.~Imine, P.~Molli, G.~Oster, and M.~Rusinowitch.
\newblock Proving correctness of transformation functions in real-time
  groupware.
\newblock In {\em Proceedings of the eighth conference on European Conference
  on Computer Supported Cooperative Work}, ECSCW'03, pages 277--293, Norwell,
  MA, USA, 2003. Kluwer Academic Publishers.

\bibitem{JayTOCHI07}
C.~Jay, M.~Glencross, and R.~Hubbold.
\newblock {Modeling the Effects of Delayed Haptic and Visual Feedback in a
  Collaborative Virtual Environment}.
\newblock {\em ACM Transactions on Computer-Human Interaction}, 14(2), August
  2007.

\bibitem{kermarrec01icecube}
A.-M. Kermarrec, A.~I.~T. Rowstron, M.~Shapiro, and P.~Druschel.
\newblock The {IceCube} approach to the reconciliation of divergent replicas.
\newblock In {\em Proceedings of the twentieth annual ACM symposium on
  Principles of distributed computing - PODC'01}, pages 210--218. ACM Press,
  2001.

\bibitem{martin09collaborative}
S.~Martin and D.~Lugiez.
\newblock Collaborative peer to peer edition: Avoiding conflicts is better than
  solving conflicts.
\newblock In H.~Weghorn and P.~T. Isa\'{\i}as, editors, {\em IADIS AC (2)},
  pages 124--128. IADIS Press, 2009.

\bibitem{martin10fixing}
S.~Martin and D.~Lugiez.
\newblock Fixing collaborative edition on typed documents.
\newblock In {\em CDVE}, 2010.

\bibitem{martin10scalable}
S.~Martin, P.~Urso, and S.~Weiss.
\newblock Scalable xml collaborative editing with undo.
\newblock In R.~Meersman, T.~Dillon, and P.~Herrero, editors, {\em On the Move
  to Meaningful Internet Systems: OTM 2010}, volume 6426 of {\em Lecture Notes
  in Computer Science}, pages 507--514. Springer, 2010.

\bibitem{oster06tombstone}
G.~Oster, P.~Urso, P.~Molli, and A.~Imine.
\newblock Tombstone transformation functions for ensuring consistency in
  collaborative editing systems.
\newblock In {\em The Second International Conference on Collaborative
  Computing: Networking, Applications and Worksharing (CollaborateCom 2006)},
  Atlanta, Georgia, USA, November 2006. IEEE Press.

\bibitem{preguica09commutative}
N.~M. Pregui\c{c}a, J.~M. Marqu{\`e}s, M.~Shapiro, and M.~Letia.
\newblock A commutative replicated data type for cooperative editing.
\newblock In {\em ICDCS}, pages 395--403. IEEE Computer Society, 2009.

\bibitem{ResselCSCW96}
M.~Ressel, D.~Nitsche-Ruhland, and R.~Gunzenh{\"a}user.
\newblock {An Integrating, Transformation-Oriented Approach to Concurrency
  Control and Undo in Group Editors}.
\newblock In {\em Proceedings of the ACM Conference on Computer-Supported
  Cooperative Work - CSCW '96}, pages 288--297, Boston, MA, USA, November 1996.
  ACM Press.

\bibitem{saito05optimistic}
Y.~Saito and M.~Shapiro.
\newblock Optimistic replication.
\newblock {\em ACM Computing Surveys}, 37(1):42--81, 2005.

\bibitem{shapiro11comprehensive}
M.~Shapiro, N.~Pregui{\c c}a, C.~Baquero, and M.~Zawirski.
\newblock {A comprehensive study of Convergent and Commutative Replicated Data
  Types}.
\newblock Research Report RR-7506, INRIA, January 2011.

\bibitem{shapiro11conflictfree}
M.~Shapiro, N.~Pregui{\c c}a, C.~Baquero, and M.~Zawirski.
\newblock Conflict-free replicated data types.
\newblock In X.~D{\'e}fago, F.~Petit, and V.~Villain, editors, {\em
  Stabilization, Safety, and Security of Distributed Systems (SSS)}, volume
  6976, pages 386--400, Grenoble, France, October 2011.

\bibitem{StCh06}
S.~Staworko and J.~Chomicki.
\newblock Validity-sensitive querying of {XML} databases.
\newblock In {\em EDBT Workshops (dataX)}, pages 164--177. Springer LNCS 4254,
  2006.

\bibitem{SuleimanGROUP97}
M.~Suleiman, M.~Cart, and J.~Ferri\'e.
\newblock {Serialization of Concurrent Operations in a Distributed
  Collaborative Environment}.
\newblock In {\em Proceedings of the ACM SIGGROUP Conference on Supporting
  Group Work - GROUP '97}, pages 435--445, Phoenix, AZ, USA, November 1997. ACM
  Press.

\bibitem{SunCSCW98}
C.~Sun and C.~Ellis.
\newblock {Operational Transformation in Real-Time Group Editors: Issues,
  Algorithms and Achievements}.
\newblock In {\em Proceedings of the ACM Conference on Computer-Supported
  Cooperative Work - CSCW '98}, pages 59--68, Seattle, WA, USA, November 1998.
  ACM Press.

\bibitem{sun98tochi}
C.~Sun, X.~Jia, Y.~Zhang, Y.~Yang, and D.~Chen.
\newblock Achieving convergence, causality preservation, and intention
  preservation in real-time cooperative editing systems.
\newblock {\em ACM Transactions on Computer-Human Interaction (TOCHI)},
  5(1):63--108, March 1998.

\bibitem{terry95bayou}
D.~B. Terry, M.~M. Theimer, K.~Petersen, A.~J. Demers, M.~J. Spreitzer, and
  C.~H. Hauser.
\newblock Managing update conflicts in {Bayou}, a weakly connected replicated
  storage system.
\newblock In {\em Proceedings of the fifteenth ACM symposium on Operating
  systems principles - SOSP'95}, pages 172--182. ACM Press, 1995.

\bibitem{weiss07wooki}
S.~Weiss, P.~Urso, and P.~Molli.
\newblock { Wooki: a P2P Wiki-based Collaborative Writing Tool}.
\newblock In {\em Web Information Systems Engineering}, pages 503--512, Nancy,
  France, December 2007. Springer.

\bibitem{weiss09logoot}
S.~Weiss, P.~Urso, and P.~Molli.
\newblock Logoot: A scalable optimistic replication algorithm for collaborative
  editing on p2p networks.
\newblock In {\em 29th IEEE International Conference on Distributed Computing
  Systems (ICDCS 2009)}, pages 404 --412, Montr\'eal, Qu\'ebec, Canada, jun.
  2009. IEEE Computer Society.

\bibitem{weiss10logootundo}
S.~Weiss, P.~Urso, and P.~Molli.
\newblock Logoot-undo: Distributed collaborative editing system on p2p
  networks.
\newblock {\em IEEE Transactions on Parallel and Distributed Systems},
  21:1162--1174, 2010.

\bibitem{Wu1995346}
S.~Wu, U.~Manber, and E.~Myers.
\newblock A subquadratic algorithm for approximate regular expression matching.
\newblock {\em Journal of Algorithms}, 19(3):346 -- 360, 1995.

\end{thebibliography}

\end{document}